\documentclass[journal,onecolumn,12pt,twoside]{IEEEtranTMBMC}

\normalsize

\usepackage[cmex10]{amsmath}
\usepackage{empheq}
\usepackage{graphicx}
\usepackage{cite}
\usepackage{xcolor}
\usepackage{siunitx}
\usepackage[nolist,nohyperlinks]{acronym}
\usepackage{booktabs}
\usepackage{amssymb}
\usepackage{amsfonts}
\usepackage{multirow}
\usepackage{rotating}
\usepackage{url}

\renewcommand{\baselinestretch}{1.4} 
\usepackage{caption}
\newcommand{\Conv}{
  \mathop{\scalebox{1.5}{\raisebox{-0.2ex}{$\circledast$}}
  }
}
\DeclareMathOperator*{\argmax}{arg\,max}	

\makeatletter
\newcommand\footnoteref[1]{\protected@xdef\@thefnmark{\ref{#1}}\@footnotemark}
\makeatother

\acrodef{1-D}[1-D]{one-dimensional}
\acrodef{3-D}[3-D]{three-dimensional}
\acrodef{MC}[MC]{synthetic molecular communication}
\acrodef{EM}[EM]{electromagnetic wave}
\acrodef{CAM}[CAM]{chorioallantoic membrane}
\acrodef{CVS}[CVS]{cardiovascular system}
\acrodef{CIR}[CIR]{channel impulse response}
\acrodef{DOI}[DOI]{digital object identifier}
\acrodef{FWHM}[FWHM]{full width at half maximum}
\acrodef{ICG}[ICG]{indocyanine green}
\acrodef{IoBNT}[IoBNT]{Internet of Bio-Nano Things}
\acrodef{ISI}[ISI]{inter-symbol-interference}
\acrodef{PDE}[PDE]{partial differential equation}
\acrodef{PDF}[PDF]{probability density function}
\acrodef{SPION}[SPION]{superparamagnetic iron-oxide nanoparticle}
\acrodef{SNR}[SNR]{signal-to-noise ratio}
\acrodef{TX}[TX]{transmitter}
\acrodef{RX}[RX]{receiver}
\acrodef{VN}[VN]{vessel network}
\acrodefplural{SPION}[SPIONs]{superparamagnetic iron-oxide nanoparticles}
\acrodefplural{LBVN}[LBVNs]{linear branched vessel networks}
\acrodefplural{SNR}[SNRs]{signal-to-noise ratios}
\acrodefplural{RX}[RXs]{receivers}
\acrodefplural{GFP}[GFPs]{green fluorescent proteins}
\acrodefplural{CIR}[CIRs]{channel impulse responses}
\acrodefplural{PDE}[PDEs]{partial differential equations}
\acrodefplural{VN}[VNs]{vessel networks}

\newtheorem{remark}{Remark}

\begin{document}

\title{Vessel Network Topology in Molecular Communication: Insights from Experiments \\and Theory\vspace*{-3mm}
\thanks{This work was funded in part by the Deutsche Forschungsgemeinschaft (DFG, German Research Foundation) -- GRK 2950 -- ProjectID 509922606, in part by the German Federal Ministry of Research, Technology and Space (BMFTR) under Project Internet of Bio-Nano-Things (IoBNT) -- grant number 16KIS1987, and in part by the Horizon Europe Marie Skodowska Curie Actions (MSCA)-UNITE under Project 101129618.}
}

\author{
\IEEEauthorblockN{Timo Jakumeit$^1$, Lukas Brand$^1$, Jens Kirchner$^2$, Robert Schober$^1$, and Sebastian Lotter$^1$}\\
\thanks{This paper was presented in part at the IEEE International Conference on Communications, 2025 \cite{Jakumeit2025}.}
\IEEEauthorblockA{\small $^1$Friedrich-Alexander-Universit\"at Erlangen-N\"urnberg, Erlangen, Germany\\
$^2$Fachhochschule Dortmund, University of Applied Sciences and Arts, Dortmund, Germany}
}

\maketitle

\begin{abstract}
The notion of \ac{MC} refers to the transmission of information via signaling molecules and is foreseen to enable innovative medical applications in the human \ac{CVS}. 
Crucially, the design of such applications requires accurate and experimentally validated channel models that characterize the propagation of signaling molecules, not just in individual blood vessels, but in complex \acp{VN}, as prevalent in the \ac{CVS}. 
However, experimentally validated models for \ac{MC} in \acp{VN} remain scarce.
To address this gap, we propose a novel channel model for \ac{MC} in complex \ac{VN} topologies, which captures molecular transport via advection, molecular and turbulent diffusion, as well as adsorption and desorption at the vessel walls.
We specialize this model for \acp{SPION} as signaling molecules by introducing a new \ac{RX} model for planar coil inductive sensors, enabling end-to-end experimental validation with a dedicated \ac{SPION} testbed.
Validation covers a range of channel topologies, from single-vessel topologies to branched \acp{VN} with multiple paths between \ac{TX} and \ac{RX}.
Additionally, to quantify how the \ac{VN} topology impacts signal quality, and inspired by multi-path propagation models in conventional wireless communications, we introduce two metrics, namely \textit{molecule delay} and \textit{multi-path spread}. We show that these metrics link the \ac{VN} structure to molecule dispersion induced by the \ac{VN} and mediately to the resulting \ac{SNR} at the \ac{RX}.
The proposed \ac{VN} structure-\ac{SNR} link is validated experimentally, demonstrating that the proposed framework can support tasks such as optimal sensor placement in the \ac{CVS} or the identification of suitable testbed topologies for specific \ac{SNR} requirements.
All experimental data are openly available on Zenodo.
\end{abstract}

\acresetall

\section{Introduction}\label{sec:Introduction}
The exchange of information between biological entities via signaling molecules lies at the core of nature's communication processes within living organisms. Over the past decade, this natural communication paradigm has been re-imagined for technological applications under the name of \ac{MC}. By encoding information in the molecule concentration, molecule type or structure, release time within a symbol interval, or in molecule mixtures~\cite{Jamali2019,Jamali2023}, \ac{MC} is envisioned to enable a wide range of innovative medical applications, most of which are intended for use inside the human body~\cite{Felicetti2016}. In this context, \ac{MC} is preferred over traditional communication schemes, such as \ac{EM}-based communication, primarily due to its inherent compatibility with biological systems.

Within the human body, the \ac{CVS} represents a particularly promising application domain for \ac{MC}. The bloodstream offers a fast and pervasive transport medium for signaling molecules, while the surrounding tissues provide direct access to key physiological processes, including metabolic regulation, immune response, and hormonal signaling~\cite{Aaronson2012}. 
For \textit{diagnostic} applications, this enables the sensing of a broad spectrum of health-related parameters and biomarkers in the bloodstream, which can be analyzed to support medical diagnoses. Examples include the early detection and localization of cancerous tissue and the detection of onset atherosclerosis\cite{Mosayebi2019,Etemadi2023,Sun2022,Khaloopour2021}.
For \textit{therapeutic} applications, \ac{MC} in the \ac{CVS} allows for the targeted delivery of drug molecules to nearly any location in the body, facilitating precise and efficient treatment strategies~\cite{Chahibi2015a,Chahibi2017,ChudeOkonkwo2017,Simo2024}. 
This can be implemented using, e.g., pH sensitive nanoparticles that selectively accumulate at diseased regions and release therapeutic agents directly into the tumor environment upon pH stimuli~\cite{Gao2010}. 
A further envisioned exemplary \textit{diagnostic and therapeutic} application is the \ac{IoBNT}, in which small-scale devices, termed \textit{gateways}, are positioned along the \ac{CVS} to receive molecular signals and convert them into \ac{EM} signals for wireless transmission to an external computational unit that performs signal processing and evaluation. These gateways are bidirectional, i.e., they can also receive \ac{EM} signals from the external unit, convert them into molecular signals, and transmit them internally for actuation. This permits the seamless integration of \ac{MC} and \ac{EM}-based communication, enabling advanced applications such as real-time health monitoring and personalized treatment~\cite{Akyildiz2015,Zafar2021,Kuscu2021}.

To support the development of such envisioned medical applications and establish a solid theoretical foundation, numerous modeling studies have examined advective-diffusive molecule transport in simplified environments, such as unbounded media, cylindrical ducts, and rectangular ducts~\cite{Schafer2021, Wicke2022, Lo2019, Kooten1996,Jiang2022,Kuscu2018,Schafer2019,Wicke2018,Noel2014a,Noel2014b,Varshney2019, Dhok2022}, see also the reviews in~\cite{Farsad2016,Jamali2019}. 
These studies address a wide range of physical and communication-related aspects, including different molecule injection methods~\cite{Schafer2021, Wicke2022}, absorption~\cite{Lo2019} and reversible sorption~\cite{Kooten1996,Jiang2022} at channel walls, ligand-receptor binding~\cite{Kuscu2018}, external influences such as gravity and magnetic fields~\cite{Schafer2019}, different molecule transport regimes where the relative influence of advection and diffusion varies~\cite{Wicke2018}, flow directions misaligned with the \ac{TX}-\ac{RX} axis~\cite{Noel2014a}, as well as (sub-)optimal sequence detection~\cite{Noel2014b} and cooperative \ac{MC} strategies~\cite{Varshney2019, Dhok2022}. 
Collectively, these works contribute to identifying key factors that govern the performance of \ac{MC} systems in simple single-vessel systems or unbounded environments. 

However, to obtain insights for future real-world applications, realistic \ac{MC} models must account for the structural complexity of the \ac{CVS}, which comprises interconnected vessels of varying shapes and sizes. In response, a growing body of research focuses on extending molecule transport models to \ac{VN} structures, aiming to capture the actual physiological conditions more accurately.
A \ac{VN} graph model capturing blood flow, pressure, and oxygen transport in \ac{3-D} brain vasculature is presented in~\cite{Reichold2009}. 
In~\cite{Chahibi2013}, a channel model based on \acp{CIR} is developed for \acp{VN}, focusing on arterial trees; 
this work is extended in~\cite{Chahibi2015,Chahibi2015a} to include adhesion and absorption at channel walls, molecule degradation, and antibody-mediated drug delivery.
A system-theoretic framework for optimally designing branched microfluidic \ac{MC} channels with rectangular cross-section is introduced in~\cite{Bicen2013}. 
The study in~\cite{Mosayebi2019} proposes a method for early cancer detection using mobile nanosensors navigating the \ac{CVS}, assuming a quasi-steady-state concentration of signaling molecules. 
Dosage estimation for cellular vaccines injected into the \ac{CVS} in the context of COVID-19 treatment is addressed in~\cite{Pal2023}. 
Microbubble propagation in \acp{VN}, including microbubble degradation effects, is modeled using a graph-based framework in~\cite{Tjabben2024}. 
Finally, the authors of~\cite{d’Esposito2018} present a model for spatiotemporally resolved drug delivery in tumors, informed by \ac{3-D} vascular imaging.
Despite these advances, a comprehensive understanding of \ac{MC} in \acp{VN} is still missing, as the existing models are either limited to local branching phenomena~\cite{Pal2023,Chahibi2013, Chahibi2015, Chahibi2015a}, are not suited for time-varying molecule injection~\cite{Mosayebi2019}, or do not fully capture the underlying physical transport principles~\cite{Tjabben2024}, and none provides a higher-level understanding of the impact that the \ac{VN} topology has on \ac{MC}.

Beyond \textit{theoretical} modeling, the feasibility of \ac{MC} has been repeatedly demonstrated \textit{experimentally} in a broad range of fluidic testbeds comprising \textit{single vessels} as communication channels.
These testbeds use diverse combinations of signaling molecules and sensor technologies, such as ions and a pH probe~\cite{Farsad2017} or electrodes~\cite{Angerbauer2023}, fluorescent nanoparticles and an optical sensor~\cite{Tuccitto2017}, glucose and an electrolyte-gated field-effect transistor~\cite{Koo2020}, \acp{SPION} and inductive or capacitive sensors~\cite{Bartunik2023, Bartunik2023a}, color pigments and optical sensors~\cite{Wietfeld2024, Pan2022}, oil and a capacitve sensor~\cite{Huang2024}, or fluorescent proteins and dyes and optical sensors~\cite{Brand2024,Lin2024}.
Recent surveys summarizing advances in fluidic MC testbeds can be found in~\cite{Lotter2023a, Hamidovic2024}.

While theoretical research on \ac{MC} in \acp{VN} is getting increasingly advanced, and experimental validation in single vessels is plentiful, experimental validation in \acp{VN} that resemble the structure and flow characteristics of the \ac{CVS} remains scarce, pointing to a critical gap in the current body of work. To the best of the authors’ knowledge, only three experimental studies pointed into this direction have been published to date.
In~\cite{Vakilipoor2025}, the first \textit{in-vivo} \ac{MC} testbed is introduced, exploiting the chicken \ac{CAM} model and its vasculature as a platform for \ac{ICG}-based \ac{MC}.
Secondly, in~\cite{Yu2024}, a \ac{3-D}-printed branched testbed with ultra-thin, semipermeable, and spatially functionalizable walls is proposed, closely mimicking important properties of the \ac{CVS}, but lacking a detailed mathematical characterization of the channel properties.
Thirdly, in~\cite{Wang2020}, a salinity-based testbed comprising a channel with two paths between \ac{TX} and \ac{RX} is used to validate an \ac{ISI}-aware decoding algorithm. 
Although the existing studies partially provide physical models for isolated sections of the channel, what remains absent in \cite{Vakilipoor2025, Yu2024, Wang2020} is a comprehensive physical channel model that links molecule propagation to the underlying network topology and explicitly captures the impact of multi-path propagation of signaling molecules in \acp{VN}.

To address the lack of a higher-level understanding of the impact that \ac{VN} topology has on \ac{MC}, and the absence of experimentally validated models for \ac{MC} in \acp{VN}, we propose an analytical end-to-end model for the transport and reception of arbitrary signaling molecules in \acp{VN}, validated through branched-channel testbed experiments using \acp{SPION} as signaling molecules. 
Based on the channel model from~\cite{Chahibi2013}, our approach incorporates additional physical effects, including turbulent diffusion at the injection site and \ac{VN} branching points, as well as adsorption and desorption at vessel walls. 
Together with advection and molecular diffusion, these effects capture the dominant transport dynamics in \acp{VN}, as confirmed experimentally. 
The channel model is coupled with a transparent \ac{RX} model and specialized to \acp{SPION} and an inductive planar coil \ac{RX}, enabling direct comparison between analytical predictions and experimental measurements. For validation, we extend the \ac{SPION}-based testbed from~\cite{Bartunik2023} to branched topologies with multiple transport paths between \ac{TX} and \ac{RX}.

Using this combined framework, we focus on quantifying how \ac{VN} topology impacts signal quality at any target location in the network. To this end, we introduce two metrics, namely \textit{molecule delay} and \textit{multi-path spread}, which span the so-called \textit{dispersion space}, linking \ac{VN} topology to molecule dispersion. The level of dispersion, in turn, directly relates to the received \ac{SNR}, allowing signal quality to be characterized \textit{solely based on topology}.
Altogether, the proposed framework provides a foundation for several practical applications. 
First, it enables the evaluation and optimization of sensor placement strategies in complex \acp{VN}, particularly under specific \ac{SNR} constraints. 
Second, it can provide a guideline for constructing experimental testbeds by predicting the structural complexity allowed to still ensure sufficient signal strength at the~\ac{RX}.
Third, the model enables a transparent quantification of how both individual path delays and the differences between them shape the overall \ac {VN} behavior. This facilitates the classification of structurally diverse \ac{VN} topologies into equivalence classes based on their dispersion characteristics and offers a powerful tool for characterizing and comparing \acp{VN}, extending beyond what existing models typically allow.
In the long term, this study aims to contribute toward a more abstract and generalizable understanding of \ac{MC} in \acp{VN}.

The main contributions of this work can be summarized as follows:
\begin{enumerate}
    \item We present a novel channel model for \ac{MC} in \acp{VN}, incorporating turbulent diffusion at the injection site and at \ac{VN} branching points, as well as molecule adsorption and desorption at the channel walls. These effects, alongside advection and molecular diffusion, are experimentally shown to significantly influence molecule transport.
    \item We specialize the generic model to \acp{SPION} as signaling molecules by deriving a novel analytical model for an inductive coil \ac{RX}.
    \item Based on the proposed model, we derive two metrics that define the \textit{dispersion space}, linking \ac{VN} topology to signal dispersion and thereby to signal quality at any target location in the network.
    \item As a practical extension of the testbed in~\cite{Bartunik2023}, we introduce branched channel topologies with two transport paths between \ac{TX} and \ac{RX}, and use this setup to validate both the end-to-end model and the dispersion space concept.
\end{enumerate}

Compared to the conference version~\cite{Jakumeit2025} of this paper, the proposed model also incorporates molecule adsorption and desorption at the channel walls as additional transport mechanisms.
Furthermore, the model in~\cite{Jakumeit2025} was validated using existing data from the \ac{SPION} testbed in~\cite{Bartunik2023}, obtained from a single-vessel channel. 
In contrast, the present study offers a more comprehensive experimental validation by additionally incorporating measurements from branched channel topologies. 
Moreover, the scope of this paper significantly extends beyond~\cite{Jakumeit2025} by providing experimental validation of the proposed dispersion space.

The remainder of this paper is organized as follows: Section~\ref{sec:System_Model} introduces the system model. 
Based on this, Section~\ref{sec:Topology_Dispersion_Metrics} derives metrics that relate \ac{VN} topology to the \ac{SNR} at the \ac{RX}.
Section~\ref{sec:Experimental_Setup} describes the experimental setup and specializes the system model to a \ac{SPION}-based end-to-end framework, enabling the comparison between model predictions and experimental results. 
Section~\ref{sec:Results} presents the experimental end-to-end validation of the model, along with numerical and experimental results for the dispersion space. 
Final conclusions and an outlook on future work are provided in Section~\ref{sec:Conclusion}.

\section{System Model}\label{sec:System_Model}
\begin{figure*}
    \centering
    \includegraphics[width=\linewidth]{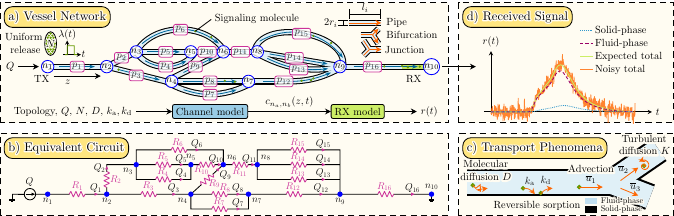}
    \caption{\textbf{System model.} \textbf{a)}~$N$ signaling molecules are uniformly released in the cross-section of the network inlet (\ac{TX}) according to injection function $\lambda (t)$, propagate through an exemplary \ac{VN} comprised of pipes, bifurcations, and junctions, and are received by a transparent \ac{RX}. \textbf{b)}~The flow rate $Q_i$ in pipe $p_i$ and the cross-sectional average flow velocity $\overline{u}_i=Q_i/(\pi r_i^2)$ are determined using an equivalent electrical circuit that models hydraulic resistance. \textbf{c)}~Molecule transport is governed by molecular and turbulent diffusion, advection, and reversible sorption at the channel walls. \textbf{d)}~Through the specialization of the generic \ac{RX} model, any transparent real-world sensor can be modeled. In any received signal, there is a contribution from molecules in fluid- and solid-phase (i.e., wall-bound).}
    \label{fig:System_Model}
\end{figure*}

In this section, we introduce the system model comprising the molecule release, i.e., the injection process at the \ac{TX}, the propagation of molecules in the channel governed by advection, diffusion, and adsorption and desorption at the channel walls, as well as a transparent \ac{RX}, cf.~Fig.~\ref{fig:System_Model}.
The system model is generically applicable to different types of signaling molecules and corresponding transparent \ac{RX} architectures.
In Section~\ref{sec:Experimental_Setup}, the system model is specialized to the use of \acp{SPION} as signaling molecules.

\subsection{Molecule Injection}\label{ssec:MoleculeInjection}
Beginning at time $t=\SI{0}{\second}$, $N$ signaling molecules are injected by the \ac{TX}, i.e., at the inlet of the network at longitudinal coordinate $z=\SI{0}{\meter}$ of the corresponding vessel, cf.~Fig.~\ref{fig:System_Model}a). 
The injection may take place over an extended period of time, according to some injection function $\lambda(t)$ given in unit $\SI{}{\per\second}$, with $\int_{0}^\infty\lambda (t)\,\mathrm{d}t=N$.
\begin{remark}
    In practical systems, the injection via a Y-connector or venous cannula induces local turbulence in the fluid flow~\cite{Wicke2022}, which results in the immediate radial dispersion of the signaling molecules.
    Further downstream in the \ac{VN}, branching points reinduce turbulences in the flow.
    Accordingly, we postulate a uniform distribution of signaling molecules throughout the pipe cross-section from the moment of injection onwards.
    Consequently, throughout this work, we employ a spatially \ac{1-D} model for molecule transport.
    The accuracy of this assumption has been previously verified in~\cite{Jakumeit2025} and is verified again by the numerical and experimental results presented in Section~\ref{sec:Results}.
\end{remark}

\subsection{Vessel Network Definition}
\acp{VN}, as found in the \ac{CVS}, generally exhibit complex topologies, comprising interconnected vessels of varying lengths, curvatures, and irregular cross-sections. 
To facilitate analysis, these networks are often approximated using simplified models for their segments~\cite{Chahibi2013,Mosayebi2019,Pal2023}. 
Accordingly, we consider \acp{VN} consisting of connected pipes, bifurcations, and junctions, as illustrated in~Fig.~\ref{fig:System_Model}a).
\begin{enumerate}
\itemsep0em
	\item \textit{Pipe:} A pipe $p_i$ is a cylindrical vessel carrying fluid from its inlet to its outlet and is defined by its length $l_i$ and radius $r_i$. Pipes can be connected to other pipes, bifurcations or junctions at both the inlet and the outlet.
	\item \textit{Bifurcation:} A bifurcation is a connection point with no spatial extent, where an inflow pipe branches into several outflow pipes. Before and after each bifurcation, there must be connected pipes.
	\item \textit{Junction:} A junction $j_m$ is a connection point with no spatial extent, where several inflow pipes merge into one outflow pipe. Before and after each junction, there must be connected pipes. We denote the set of all inflow pipes of junction $j_m$ by $\mathcal{I}(j_m)$.
\end{enumerate}
Subsequently, we refer to the network inlet, outlet, and connection points, i.e., bifurcations and junctions, as nodes $n_v$, $\, v\in \left\lbrace 1,\ldots , V\right\rbrace $, where $V$ is the total number of nodes, and $n_1$ and $n_V$ are the nodes at the network inlet and outlet, respectively.
Pipes correspond to directed edges between these nodes, with the direction of the edge determined by the flow direction. 
This allows for a simplified representation of any \ac{VN} as a directed graph, as depicted with blue nodes and dark blue arrows in Fig.~\ref{fig:System_Model}a) for an exemplary topology.
In addition, $\mathcal{P}(n_a,n_b)$ denotes the set of all paths between nodes $n_a$ and $n_b$ that differ in at least one pipe, with $a,b\in \left\lbrace 1,\ldots , V \right\rbrace$. 
Here, a path $P_k$ comprises a subset of connected pipes and junctions of the network and is denoted as
\begin{equation}\label{eqn:PathSet}
P_k=\left\lbrace p_i\,|\, i\in \mathcal{E}_k  \right\rbrace \cup \left\lbrace j_m\,|\, m\in \mathcal{J}_k \right\rbrace\,\text{,}
\end{equation}
where $\mathcal{E}_k\subseteq \left\lbrace 1,\ldots ,E \right\rbrace $ and $\mathcal{J}_k\subseteq \left\lbrace 1,\ldots ,J \right\rbrace $ are the index sets of the pipes and junctions in $P_k$, respectively, and $E$ and $J$ denote the number of pipes and junctions in the \ac{VN}, respectively.
Bifurcations are implicitly accounted for in $\mathcal{P}(n_a,n_b)$, as they determine which paths exist. In this work, we focus on \acp{VN} comprising exactly one inlet and one outlet.

\subsection{Fluid Flow in the Vessel Network}
Applying a flow rate $Q$ at the network inlet induces a pressure gradient between $n_1$ and $n_V$, driving fluid flow throughout the \ac{VN}, with zero pressure assumed at the outlet $n_V$. The flow rate $Q_i$ in unit $\SI{}{\meter\cubed\per\second}$ in pipe $p_i$ is obtained by assigning a hydraulic resistance~\cite[Eq.~(14)]{Chahibi2013}
\begin{equation}
R_i = \dfrac{8\mu l_i}{\pi r_i^4}
\end{equation}
to each pipe and constructing an equivalent electrical circuit matching the \ac{VN} topology with electrical ground at $n_V$, cf.~Fig.~\ref{fig:System_Model}b). Here, $\mu$ denotes the dynamic fluid viscosity. Modeling $Q$ as a current source and solving the circuit for its currents yields the individual flow rates $Q_i$~\cite{Chahibi2013}, from which the average cross-sectional velocities in unit $\SI{}{\meter\per\second}$ follow as
\begin{equation}
\overline{u}_i = \dfrac{Q_i}{\pi r_i^2}\,\text{.}
\end{equation}

\subsection{Effective Diffusion}
Molecules in \acp{VN} disperse due to multiple types of diffusion. Firstly, thermal vibrations in the fluid induce molecular diffusion, characterized by the molecular diffusion coefficient~\cite[Eq.~(44)]{Chahibi2013}
\begin{equation}\label{eqn:MolecularDiffusionCoefficient}
	D=\dfrac{k_\mathrm{B}T}{6\pi\mu R}\,\text{,}
\end{equation}
where $k_\mathrm{B}=\SI{1.38e-23}{\meter\squared\kilo\gram\per\second\squared\per\kelvin}$, $T$, and $R$ denote the Boltzmann constant, the fluid temperature in unit $\SI{}{\kelvin}$, and the signaling molecule radius in unit $\SI{}{\meter}$, respectively. Secondly, turbulent mixing at the point of injection and at branching points in the \ac{VN} induces turbulent diffusion. While modeling individual turbulences is analytically intractable, a simple way to represent this type of molecular transport is by assuming underlying laminar flow in conjunction with eddy diffusion, represented by the eddy diffusion coefficient~\cite[Eq.~(4.16)]{Nieuwstadt2016}
\begin{equation}\label{eqn:TurbulentDiffusionCoefficient}
	K_i=\alpha \overline{u}_i r_i,\quad \alpha\in [0,2]\,\text{.}
\end{equation}
Here, $\alpha$ denotes a unitless proportionality constant.
In a given pipe $p_i$, $K_i$ is proportional to the distance orthogonal to the flow direction over which eddies persist, which is proportional to the pipe radius $r_i$, and flow velocity $\overline{u}_i$~\cite{Nieuwstadt2016}. 
Molecular and eddy diffusion can be modeled as additive phenomena and are both incorporated in the apparent diffusion coefficient $\overline{D}_i=D+K_i$.
The shear-induced dispersion from laminar flow combined with the apparent diffusion lead to an \textit{effective} dispersion, described by the Aris-Taylor effective diffusion coefficient~\cite[Eq.~(26)]{Aris1956}
\begin{equation}\label{eqn:ArisTaylorEffectiveDiffusionCoefficient}
	D^\mathrm{eff}_i=\dfrac{r_i^2 \overline{u}_i^2}{48 \overline{D}_i}+\overline{D}_i\,\text{.}
\end{equation}

\subsection{Molecule Transport in a Single Vessel}\label{ssec:MoleculeTransport}

Below, we first model molecule transport in a single pipe, before extending the model to \acp{VN} in Subsection~\ref{ssec:ChannelModelVesselNetwork}. The molecule transport is governed by advection, diffusion, and adsorption and desorption at the channel walls. The latter two phenomena are jointly referred to as \textit{sorption}.
The transport phenomena are illustrated in Fig.~\ref{fig:System_Model}c).

Within a pipe, molecules in the \textit{fluid phase}, i.e., dissolved in the fluid, propagate via advection and diffusion. Along their way from the inlet to the outlet, they may repeatedly adsorb to and desorb from the pipe wall. Adsorbed molecules reside in the so-called \textit{solid phase} and are immobile. 
Assuming first-order sorption kinetics that are uniform along the pipe wall, the fluid-phase concentration $c_i(z,t)$ and solid-phase concentration $s_i(z,t)$ of the molecules in pipe $p_i$ obey a two-compartment model that captures advective, diffusive, and sorptive dynamics~\cite[Eqs.~(1a), (1b)]{Kooten1996}
\begin{subequations}\label{eqn:CompartmentModel}
    \begin{align}
        \begin{split}
            \frac{\partial c_i(z,t)}{\partial t} 
            &=D_i^{\mathrm{eff}} \frac{\partial^2 c_i(z,t)}{\partial z^2}
            - \overline{u}_i \frac{\partial c_i(z,t)}{\partial z}- \frac{\partial s_i(z,t)}{\partial t}\,\text{,}
        \end{split}
        \label{eqn:CompartmentModelTransport} \\[0.5ex]
        \frac{\partial s_i(z,t)}{\partial t} 
        &= k_{\mathrm{a}} c_i(z,t) - k_{\mathrm{d}} s_i(z,t)\label{eqn:CompartmentModelSorption}
        \,\text{.}
    \end{align}
\end{subequations}
Here, $z$ denotes the longitudinal coordinate \textit{within $p_i$}, i.e., reaching from its inlet at $z=0$ to its outlet at $z=l_i$. $D^\mathrm{eff}_i$ denotes the effective diffusion coefficient in $p_i$ in unit $\SI{}{\meter\squared\per\second}$. $k_\mathrm{a}$ and $k_\mathrm{d}$ denote the adsorption and desorption rate constants in units $\SI{}{\per\second}$, respectively, cf.~Fig.~\ref{fig:System_Model}c).
These constants are assumed to be space-independent, which is an accurate approximation in vessels with small diameters and strong cross-sectional mixing, where the cross-sectional position has little effect on the adsorption probability, as further confirmed by the numerical results in Section~\ref{sec:Results}.
To solve~\eqref{eqn:CompartmentModel}, we assume for the moment an instantaneous release of $N$ signaling molecules at $n_1$ at $t=0$. Further, we assume that initially no signaling molecules are in solid phase and $N$ molecules are present in the fluid phase, yielding the initial conditions
\begin{subequations}\label{eqn:InitialConditions}
    \begin{align}
        c_i(z,0)&= N\delta (z)\label{eqn:InitialFluidConcentration}\,\text{,}\\
        s_i(z,0)&=0\label{eqn:InitialWallConcentration}\,\text{.}
    \end{align}
\end{subequations}
Here, $\delta (\cdot)$ denotes the Dirac delta function.
A detailed solution of~\eqref{eqn:CompartmentModel}, covering the initial conditions in~\eqref{eqn:InitialConditions}, is given in~\cite{Kooten1996,Giddings1955}. 
For brevity, we provide a compact derivation based on physical intuition.

During the time that molecules reside in the fluid phase, their motion is governed by advection and diffusion, resulting in the fluid-phase concentration in unit $\SI{}{\per\meter}$~\cite[Eq.~(49)]{Chahibi2013}
\begin{equation}\label{eqn:AdvectionDiffusionFluidPhaseConcentration}
	\tilde{c}_i(z,t)\hspace*{-.5mm}=\hspace*{-.5mm}\dfrac{N}{\sqrt{4\pi D^\mathrm{eff}_it}}\exp\hspace*{-.5mm} \left(\hspace*{-1mm}-\dfrac{(z-\overline{u}_it)^2}{4D^\mathrm{eff}_it}\hspace*{-.5mm} \right)\hspace*{-1mm},\,z\hspace*{-1mm}\in\hspace*{-1mm} [0,l_i]\,\text{.}
\end{equation}
The remaining time, molecules are sorbed to the channel walls and immobile. Over a time interval $[0, t]$, the residence time $\tau$, i.e., the cumulative time a molecule spends in the fluid-phase, is stochastic and governed by the sorption kinetics. In particular, $\tau$ is distributed according to the sorption kernels $g_\mathrm{ff}(\tau ,t)$ and $g_\mathrm{fs}(\tau ,t)$ as follows~\cite[Eqs.~(2), (3)]{Kooten1996}, \cite[Eq.~(5)]{Giddings1955}
\begin{align}
    \label{eqn:sorption_kernel_ff}
    g_\mathrm{ff}(\tau ,t) &= \sqrt{\dfrac{k_\mathrm{a}k_\mathrm{d}\tau}{t-\tau}} \exp\left( - (k_\mathrm{a}-k_\mathrm{d})\tau-k_\mathrm{d}t\right) I_1\left(2\sqrt{k_\mathrm{a}k_\mathrm{d}\tau (t-\tau)} \right)\,, \\
    \label{eqn:sorption_kernel_fs}
    g_\mathrm{fs}(\tau ,t) &= k_\mathrm{a}\exp (-(k_\mathrm{a}-k_\mathrm{d})\tau - k_\mathrm{d}t) I_0\left(2\sqrt{k_\mathrm{a}k_\mathrm{d}\tau (t-\tau)} \right)\,\text{,}
\end{align}
where $I_0(\cdot)$ and $I_1(\cdot)$ are the zero- and first-order modified Bessel functions of the first kind, respectively. Here, $g_\mathrm{ff}(\tau, t)$ holds for molecules that are initially in fluid-phase, adsorb at least once, and are again in the fluid-phase at time $t$. Similarly, $g_\mathrm{fs}(\tau, t)$ holds for molecules that are initially in fluid-phase, adsorb at least once, and are in the solid-phase at time $t$. 

\begin{remark}
    We note that $g_\mathrm{ff}(\tau, t)$ is the \textit{joint probability} for a molecule spending $\tau$ time in the fluid phase during the interval $[0,t]$ \textit{and} being initially and ultimately in fluid-phase. Therefore, the equilibrium probability of a molecule being in fluid-phase, $k_\mathrm{d}/(k_\mathrm{a}+k_\mathrm{d})$, is inherently incorporated in $g_\mathrm{ff}(\tau, t)$. The same holds for $g_\mathrm{fs}(\tau, t)$ and probability $k_\mathrm{a}/(k_\mathrm{a}+k_\mathrm{d})$. The corresponding \textit{conditional probabilities} thus have the following integral properties: $\int_0^\infty g_\mathrm{ff}(\tau, t)(k_\mathrm{a}+k_\mathrm{d})/k_\mathrm{d}\,\mathrm{d}\tau=\int_0^\infty g_\mathrm{fs}(\tau, t)(k_\mathrm{a}+k_\mathrm{d})/k_\mathrm{a}\,\mathrm{d}\tau=1$.
\end{remark}

Given that a molecule remains continuously in the fluid phase over $[0,t]$ with cumulative probability $\exp(-k_\mathrm{a}t)$~\cite{Kooten1996}, the final fluid-phase concentration in unit $\SI{}{\per\meter}$ follows as~\cite[Eq.~(9a)]{Kooten1996}
    \begin{equation}\label{eqn:FluidPhaseConcentration}
        c_i(z,t) = \tilde{c}_i(z,t) \exp \left(-k_\mathrm{a} t \right) + \int\limits_0^t\tilde{c}_i(z,\tau)g_\mathrm{ff}(\tau, t)\,\mathrm{d}\tau\,\text{,}
    \end{equation}
where the first additive term on the right-hand side of~\eqref{eqn:FluidPhaseConcentration} accounts for molecules that continuously remain in the fluid phase. The second additive term captures the contribution of molecules that adsorb to the channel wall at least once in the considered time interval by marginalizing over the residence time $\tau$, governed by $g_\mathrm{ff}(\tau ,t)$. The solid-phase concentration in unit $\SI{}{\per\meter}$ is obtained as~\cite[Eq.~(9b)]{Kooten1996}
\begin{equation}\label{eqn:SolidPhaseConcentration}
        s_i(z,t) = \int\limits_0^t \tilde{c}_i(t,\tau)g_\mathrm{fs}(\tau, t)\,\mathrm{d}\tau\,\text{.}
    \end{equation}
    
\begin{figure}
    \centering
    \includegraphics[width=.75\textwidth]{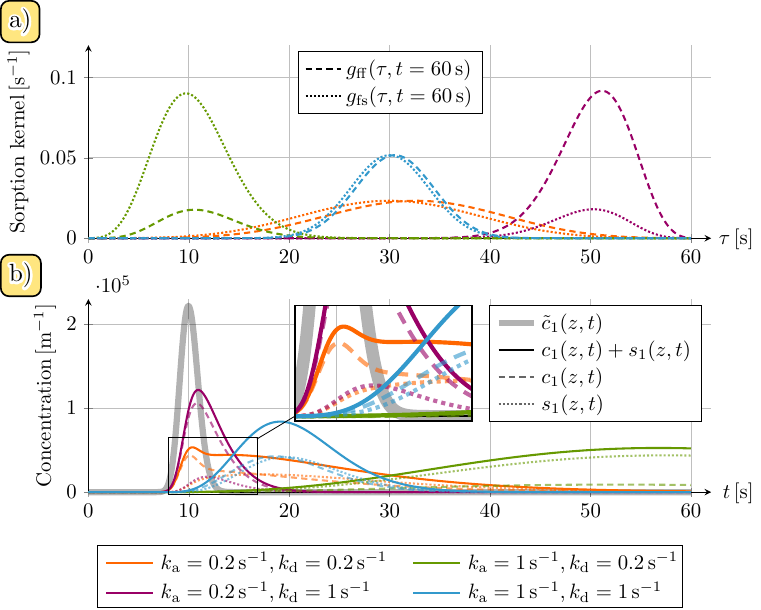}
    \caption{\textbf{Channel model behavior.} \textbf{a)} Sorption kernels and \textbf{b)} resulting concentrations for various sorption rate constants in a single pipe $p_1$. The advection-diffusion-driven concentration $\tilde{c}_1(z,t)$ is compared to the advection-diffusion-sorption-driven fluid-phase concentration $c_1(z,t)$, the solid-phase concentration $s_1(z,t)$, and the total concentration in fluid- and solid-phase $c_1(z,t)+s_1(z,t)$. Here, $z=\SI{0.2}{\meter}$, $r_1=\SI{7.95e-4}{\meter}$, $\overline{u}_1=\SI{0.02}{\meter\per\second}$, $N=\SI{1e4}{}$, $\alpha =1$, and $D=\SI{1.12e-11}{\meter\squared\per\second}$ are used.}
    \label{fig:ChannelModelBehavior}
\end{figure}
Fig.~\ref{fig:ChannelModelBehavior}a) shows the sorption kernels as defined in \eqref{eqn:sorption_kernel_ff} and \eqref{eqn:sorption_kernel_fs}, illustrated by dashed and dotted lines, respectively, for a single pipe $p_1$ at time $t=\SI{60}{\second}$. The sorption rate constants are varied to illustrate their effect on $g_\mathrm{ff}(\tau,t)$ and $g_\mathrm{fs}(\tau,t)$, respectively.
Fig.~\ref{fig:ChannelModelBehavior}b) depicts the resulting fluid-, solid-, and combined-phase concentration profiles for different sorption rate constants, indicated by different colors. For reference, the advection-diffusion-only solution, i.e., for $k_\mathrm{a}=0$, is shown as a thick gray line. The balance between fluid- and solid-phase contributions, and thus the overall concentration shape and delay, varies significantly with $k_\mathrm{a}$ and $k_\mathrm{d}$, cf.~Fig.~\ref{fig:ChannelModelBehavior}b). The reason for this becomes apparent when considering, e.g., the purple curves in~Fig.~\ref{fig:ChannelModelBehavior}a). For small $k_\mathrm{a}$ and large $k_\mathrm{d}$, in the time window from $t=\SI{0}{\second}$ to $t=\SI{60}{\second}$, molecules are likely to spend most of the time in the fluid phase, i.e., $g_\mathrm{ff}(\tau,\SI{60}{\second})$ and $g_\mathrm{fs}(\tau,\SI{60}{\second})$ are distributed around large $\tau$. 
Moreover, since $k_\mathrm{a}\ll k_\mathrm{d}$, at $t=\SI{60}{\second}$, molecules are much more likely to be in the fluid phase, compared to the solid phase, which is apparent when comparing the amplitudes of the dashed and dotted purple curves in Fig.~\ref{fig:ChannelModelBehavior}a). 
In contrast, for strong $k_\mathrm{a}$ and low $k_\mathrm{d}$, i.e., the green curves in Fig.~\ref{fig:ChannelModelBehavior}a), molecules spend most time in the solid phase (and consequently little time in the fluid phase), and their sorption kernels are thus centered around smaller $\tau$. Furthermore, at $t=\SI{60}{\second}$, molecules are much more likely to be in the solid phase, compared to the fluid phase, which is apparent when comparing the amplitudes of the dashed and dotted green curves in Fig.~\ref{fig:ChannelModelBehavior}a).
The orange and blue curves in Fig.~\ref{fig:ChannelModelBehavior}a) and \ref{fig:ChannelModelBehavior}b) represent intermediate cases in which $k_\mathrm{a}$ and $k_\mathrm{d}$ are equal.

Since molecules are mobile only while in the fluid phase, the instantaneous net molecule flux in pipe $p_i$ normalized to a single molecule in unit $\SI{}{\per\second}$ is derived from~\eqref{eqn:FluidPhaseConcentration} as~\cite[Eq.~(4.4)]{Berg1993}
    \begin{equation}\label{eqn:PipeFlux}
        \hspace*{-1mm}J_i(z,t) \hspace*{-.5mm}= \hspace*{-.5mm}\dfrac{1}{N}\left[ \left( \frac{z - \overline{u}_i t}{2t} + \overline{u}_i \right) \tilde{c}_i(z,t) \exp\left(-k_\mathrm{a} t\right)
        + \hspace*{-1mm}\int\limits_0^t \hspace*{-1mm}\left( \frac{z - \overline{u}_i \tau}{2\tau} + \overline{u}_i \right) \tilde{c}_i(z,\tau) g_\mathrm{ff}(\tau, t) \, \mathrm{d}\tau\right]\text{.}
    \end{equation}
We refer to Appendix~\ref{sec:AppA} for a detailed derivation of~\eqref{eqn:PipeFlux}.
Note that when the advective flux dominates the diffusive flux, as is generally the case in arteries of the human \ac{CVS}, and either $k_\mathrm{a} = 0$, or both $k_\mathrm{a} > 0$ and $k_\mathrm{d} > 0$, $J_i(z,t)$ constitutes a \ac{PDF} over time, satisfying $\int_0^\infty J_i(z,t)\,\mathrm{d} t = 1$.

\begin{remark}
    Setting $k_\mathrm{a}=0$ prohibits adsorption at the channel walls, yielding $g_\mathrm{ff}(\tau,t)=g_\mathrm{fs}(\tau,t)=0$, such that~\eqref{eqn:SolidPhaseConcentration} vanishes and~\eqref{eqn:FluidPhaseConcentration} reduces to the advection-diffusion expression in~\eqref{eqn:AdvectionDiffusionFluidPhaseConcentration}. Similarly, the flux in~\eqref{eqn:PipeFlux} reduces to the advective-diffusive flux in~\cite[Eq.~(6)]{Jakumeit2025}.
\end{remark}

\subsection{Molecule Transport in a Vessel Network}\label{ssec:ChannelModelVesselNetwork}

To characterize molecule transport in \acp{VN}, in addition to the fluid- and solid-phase concentrations and the molecule flux in a single pipe, the impact of a junction $j_m\in P_k$ that is traversed by a molecule through $p_i\in P_k$ with $p_i\in\mathcal{I}(j_m)$ is described as a multiplicative constant $\gamma_m^i$ using the principle of mass conservation~\cite[Eq.~(22)]{Mosayebi2019}
\begin{equation}\label{eqn:JunctionCIR}
	\gamma^i_m=\dfrac{Q_{i}}{\sum_{p_v\in\mathcal{I}(j_m)}Q_v},\quad 0< \gamma_m^i\leq 1\,\text{,}
\end{equation}
i.e., molecules partition according to the ratio of flow rates. 

Under the assumption of linear molecule transport, the end-to-end \ac{CIR} $h_{n_a,n_b}(z,t)$ in unit $\SI{}{\per\meter}$ of the network between two nodes $n_a$ and $n_b$ equals the sum of the individual path \acp{CIR}, which in turn result from the convolution of the molecule flux entering the \ac{RX} pipe $p_{i'}$  and the combined fluid- and solid-phase concentrations in $p_{i'}$
\begin{equation}\label{eqn:EndToEndCIR}
		h_{n_a,n_b}(z,t)= \left[ \sum\limits_{P_k\in\mathcal{P}(n_a,n_b)} \left(\Conv_{ \substack{i\in \mathcal{E}_k:\\ p_i\neq p_{i'}}} \hspace*{-2.5mm} J_i(l_i,t)\hspace*{2mm} \cdot \hspace*{-4.5mm}\prod_{\substack{(i,m)\in \mathcal{E}_k\times \mathcal{J}_k:\\ p_i\in \mathcal{I}(j_m) }} \hspace*{-7mm} \gamma^i_m \hspace*{1mm} \right)  \right] * \dfrac{1}{N}(c_{i'}(z,t)+s_{i'}(z,t))\,\text{,}
\end{equation}
where $*$ and $\Conv_{i\in\mathcal{S}\subset\mathbb{N}}J_i=J_{i_1}*J_{i_2}*\ldots *J_{i_{\vert \mathcal{S}\vert}}$ denote the convolution operator and the convolution operator over a set with respect to time, respectively. Here, $\mathbb{N}$ and $\vert \cdot \vert$ denote the set of natural numbers and the cardinality of a set, respectively.
Since molecule concentration remains unchanged across bifurcations~\cite[Eq.~(53)]{Chahibi2013}, their \acp{CIR} do not appear in~\eqref{eqn:EndToEndCIR}. From~\eqref{eqn:EndToEndCIR}, the end-to-end molecule concentration in unit $\SI{}{\per\meter}$ after the injection of molecules according to $\lambda(t)$ at $n_a $ is obtained as
\begin{equation}
    c_{n_a,n_b}(z,t)=\lambda (t) * h_{n_a,n_b}(z,t)\,\text{.}
\end{equation}

\subsection{Molecule Reception}
We consider a transparent \ac{RX} whose domain extends over $z$. The received signal results from the end-to-end concentration~as 
\begin{equation}\label{eqn:ReceivedSignal}
	r(t)=f\left( \int_{z\in\mathrm{dom}(w(\cdot))} w(z)c_{n_a,n_b}(z,t)\,\mathrm{d}z\right) \,\text{,}
\end{equation}
where $\mathrm{dom}(\cdot)$ denotes the domain operator. 
Here, the weighting function $w(z)$ and the signal conversion function $f(x)$ are specific to the sensor employed as \ac{RX}. 
Thereby, potential inhomogeneities of the sensing process\footnote{Consider, e.g., optical \acp{RX} in the \ac{CVS} where heterogeneous refractive indices of surrounding tissues cause sensing inhomogeneities over space.} over $z$ are captured by $w(z)$ and subsequent processing steps\footnote{Consider, e.g., differential \acp{RX} where the current measured signal amplitude is compared to that measured in previous time steps or a reference signal.} are modeled by $f(x)$.

In the \ac{MC} literature, often the special case of a so-called perfect counting \ac{RX}, 
with length $l_\mathrm{RX}$ centered at $z_\mathrm{RX}$ with $w(z)=1$, $\forall z\in [z_\mathrm{RX}-l_\mathrm{RX}/2,z_\mathrm{RX}+l_\mathrm{RX}/2]$, and linear mapping, i.e., $f(x)=x$, is assumed. 
In this case, the received signal equals the number of observed molecules $N^\mathrm{obs}(t)=\int_{z_\mathrm{RX}-l_\mathrm{RX}/2}^{z_\mathrm{RX}+l_\mathrm{RX}/2}c_{n_a,n_b}(z,t)\,\mathrm{d}z$.
In practice, however, sensors with such properties do not exist, emphasizing the practical importance of the sensor-related degrees of freedom in~\eqref{eqn:ReceivedSignal}.

\section{Topology-Dispersion Metrics}\label{sec:Topology_Dispersion_Metrics}
To characterize how the \ac{VN} topology affects the received molecular signal, we propose two metrics that relate the topology of \acp{VN} to the received \ac{SNR}. 
In particular, these metrics are derived from the system model in Section~\ref{sec:System_Model} and characterize how strongly molecules are dispersed as they travel through a given network topology, taking into account molecular and eddy diffusion, advection and sorption, as well as multi-path propagation. 
The degree of dispersion is in turn associated with the received \ac{SNR}. 
This methodology is akin to the characterization of multi-path channels in wireless communications, where similar metrics are employed to assess the influence of the channel on the quality of the received signal~\cite{Rappaport2024}.

\textit{1) Molecule Delay:} 
The extent of molecule dispersion via diffusion from the network inlet to the outlet is inherently linked to the molecule travel time, with longer delays implying larger dispersion, cf.~\eqref{eqn:AdvectionDiffusionFluidPhaseConcentration}.
This is captured by the \textit{molecule delay} metric, derived below.

For a single pipe $p_i$, the time when most molecules are located at its outlet, denoted as the pipe peak time, is given as
\begin{equation}\label{eqn:PipeTmax}
	t^\mathrm{peak}_{p_i}=\argmax_t \left(c_i(l_i,t)+s_i(l_i,t)\right)\,\text{.}
\end{equation}
Finding a closed-form solution for~\eqref{eqn:PipeTmax} is infeasible due to the integrals and Bessel functions involved in~\eqref{eqn:FluidPhaseConcentration} and~\eqref{eqn:SolidPhaseConcentration}. 
However, in practice, it often holds that $k_\mathrm{a}\ll k_\mathrm{d}$, cf.~Section~\ref{sec:Results}, in which case the peak time is determined mostly by advective-diffusive molecule transport\footnote{Even though the \textit{peak time} is most strongly impacted by advective-diffusive transport for $k_\mathrm{a}\ll k_\mathrm{d}$, sorption dynamics still play a significant role for the \textit{tail behavior} of received signals and can thus not be neglected in~\eqref{eqn:FluidPhaseConcentration} and~\eqref{eqn:SolidPhaseConcentration}. 
This is also apparent in Fig.~\ref{fig:Advective_Diffusive_vs_Sorptive_Transport}.}. Under this assumption, $t^\mathrm{peak}_{p_i}$ can be approximated with the closed-form expression~\cite[Eq.~(18)]{Jakumeit2025}
\begin{equation}\label{eqn:PipeTmaxApproximation}
	t^\mathrm{peak}_{p_i}\overset{k_\mathrm{a}\ll k_\mathrm{d}}{\approx} \dfrac{-D^\mathrm{eff}_i+\sqrt{{D^\mathrm{eff}_i}^2+\overline{u}_i^2l_i^2}}{\overline{u}_i^2}\,\text{.}
\end{equation}
From~\eqref{eqn:PipeTmax}, or similarly~\eqref{eqn:PipeTmaxApproximation}, the time when most molecules are located at the outlet of a path $P_k$, denoted as the path peak time, follows~as
\begin{equation}\label{eqn:PathTmax}
	t^\mathrm{peak}_{P_k}=\sum\limits_{p_i\in P_k} t^\mathrm{peak}_{p_i}\,\text{.}
\end{equation}
Moreover, using~\eqref{eqn:JunctionCIR}, the fraction of molecules traveling through $P_k$ is
\begin{equation}\label{eqn:PathGamma}
	\gamma_{P_k} = \prod\limits_{\substack{p_i,j_m\in P_k\\ p_i\in\mathcal{I}(j_m )}} \gamma^i_m,\quad 0< \gamma_{P_k}\leq 1\,\text{,}
\end{equation}
i.e., the product of the flow rate fractions of all junctions contained in the path. 
From~\eqref{eqn:PipeTmax}--\eqref{eqn:PathGamma}, we define the \textit{molecule delay} for a given network between nodes $n_a$ and $n_b$ as
\begin{equation}\label{eqn:NetworkAverageTimeOfArrivalWeightedAverage}
	t_{n_a}^{n_b}=\sum\limits_{P_k\in \mathcal{P}(n_a,n_b)} \gamma_{P_k}t^\mathrm{peak}_{P_k}\,\text{,}
\end{equation}
giving, via $\gamma_{P_k}$, more weight to paths transporting more molecules.
Note that $t_{n_a}^{n_b}$ resembles the \textit{excess delay} metric in multi-path wireless communications, quantifying the mean signal delay~\cite{Rappaport2024}.

\textit{2) Multi-Path Spread:}
Networks with paths having widely varying peak times $t^\mathrm{peak}_{P_k}$ incur increased dispersion, causing molecules to arrive at the \ac{RX} more spread out over time.
We quantify this phenomenon via the \textit{multi-path spread}
\begin{equation}\label{eqn:PTV}
	\sigma_{n_a}^{n_b}=\sqrt{\sum_{P_k\in\mathcal{P}(n_a,n_b)} \gamma_{P_k} \left(t^\mathrm{peak}_{P_k}-t_{n_a}^{n_b} \right)^2}\,\text{.}
\end{equation}
Note that $\sigma_{n_a}^{n_b}$ parallels the \textit{root mean square delay spread} metric, describing signal spread in a multi-path wireless channel~\cite{Rappaport2024}.

\begin{figure}
    \centering
    \includegraphics[width=.75\textwidth]{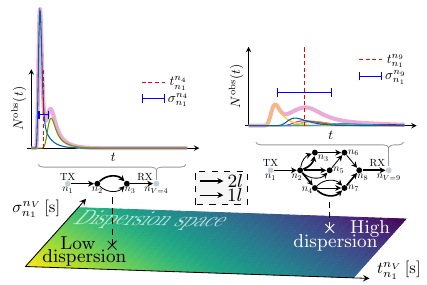}
    \caption{\textbf{Dispersion space.} The position of any \ac{VN} in the space is solely based on its topology, as $t_{n_1}^{n_V}$ and $\sigma_{n_1}^{n_V}$ characterize the dispersion of the molecules propagating from \ac{TX} to \ac{RX}. The colors of the space are indicative of the hypothesized received \ac{SNR}. Two exemplary \acp{VN} with identical pipe radii and pipe lengths of either $l$ or $2l$, along with their received signals $N^\mathrm{obs}(t)$, are shown. Thin colored curves show individual path contributions; the pink curve shows the total signal resulting from the superposition of the path contributions.}
    \label{fig:Dispersion_Space_Concept}
\end{figure}

\textit{3) Dispersion Space:}
$t_{n_a}^{n_b}$ and $\sigma_{n_a}^{n_b}$ span a two-dimensional space denoted as \textit{dispersion space}, cf.~Fig.~\ref{fig:Dispersion_Space_Concept}.
Any \ac{VN} can be located in this space \textit{solely based on its topology}.
Setting $n_a=n_1$, $n_b=n_V$, the position in the space can then be used to infer the degree to which molecules disperse while propagating from \ac{TX} to \ac{RX}.
\acp{VN} close to the origin experience comparatively little dispersion and \acp{VN} far away from the origin suffer from strong dispersion, cf.~left and right \ac{VN} in Fig.~\ref{fig:Dispersion_Space_Concept}, respectively. 
Lastly, we hypothesize that the level of dispersion correlates negatively with the received \ac{SNR} (cf.~\eqref{eqn:SNR}), as greater dispersion causes molecules to arrive more spread out in time, reducing their instantaneous concentration at the \ac{RX} and thereby lowering the signal strength relative to noise. Note that the dispersion metrics in~\eqref{eqn:NetworkAverageTimeOfArrivalWeightedAverage} and~\eqref{eqn:PTV} are computed based on \textit{concentration} signals, which generally serve as a reliable proxy for the actual sensor outputs, since most \ac{MC} sensors are explicitly designed to reflect local molecule concentrations.

\begin{remark}
    Analogous to multi-path wireless communications, where environments such as hilly terrain, urban, suburban, or indoor settings lead to characteristic received signal properties due to specific scattering effects~\cite{Rappaport2024}, the metrics introduced here for \ac{MC} in \acp{VN} provide a means to associate particular \ac{VN} types, e.g., specific organs or tissue regions, with their characteristic received molecular signal properties. Looking ahead, this may facilitate a more general understanding of \ac{MC} in specific \ac{VN} types.
\end{remark}

\section{Experimental Setup and Inductive Receiver Model}\label{sec:Experimental_Setup}
To experimentally validate the theoretical predictions of the proposed model, we specialize the channel model in Section~\ref{sec:System_Model} to \ac{SPION}-based \ac{MC}. Specifically, we employ an improved version of the \ac{SPION} testbed from~\cite{Bartunik2023}, with an upgraded background flow pump and more complex channel topologies, as an experimental benchmark and derive a novel analytical model for the planar coil inductive \ac{RX} used in the testbed, enabling comparisons between model \acp{CIR} and measured \acp{CIR}. Below, we first describe the testbed setup and measurement procedure, then we present the inductive \ac{RX} model to complete the \ac{SPION}-specialized end-to-end model.

\subsection{Branched SPION-Testbed}
    An overview of the testbed components is given in Fig.~\ref{fig:TestbedSetup}.

\subsubsection{Signaling Molecules}
\acp{SPION} are superparamagnetic nanoparticles, e.g., synthesized through co-precipitation~\cite{Palanisamy2019}. 
At their core, they hold a cluster of iron-oxide molecules, responsible for their high magnetic susceptibility $\chi_\mathrm{ref}$. 
As a result, \acp{SPION} become magnetized in the presence of an external magnetic field, but show no remanence once removed from the field, due to their small size~\cite{Bartunik2023}.
Various coatings can be applied to the core~\cite{Palanisamy2019}, including biocompatible coatings like human serum albumin, chemical degradation and agglomeration mitigating coatings like Dextran, therapeutic coatings carrying drug particles, or ligands that bind to desired target sites.
Depending on the coating, the diameters of \acp{SPION} range from below $\SI{10}{\nano\meter}$ to micrometers~\cite{Palanisamy2019}.
In the following, we used \acp{SPION} with a lauric acid coating for increased chemical stability and biocompatibility exhibiting a hydrodynamic radius of $R\approx \SI{19.2}{\nano\meter}$ and a susceptibility of $\chi_\mathrm{ref}\approx \SI{5.99e-3}{}$ (SI units).

\subsubsection{Background Flow} Distilled water at room temperature ($\approx \SI{23}{\degreeCelsius}$) was stored in a plastic cup reservoir. A time-invariant background flow of flow rate $Q=\SI{2.861e-7}{\meter\cubed\per\second}$ was established using an \textit{Ismatec Reglo-Z Analog} gear pump (ISM895-230), which drew distilled water from the reservoir and directed it into the communication channel, see Fig.~\ref{fig:TestbedSetup}a).

\subsubsection{Molecule Injection} To characterize the channel, \acp{SPION} at room temperature were injected into the channel using a \textit{Vasofix Safety Gauge 24} venous cannula, which had an inner diameter of $\SI{0.7}{\milli\meter}$ and a length of $\SI{19}{\milli\meter}$. 
The cannula featured a flexible plastic tip and was inserted into the tubing material in the downstream direction. 
The tip of the needle was considered the \ac{TX}. 
The injected \ac{SPION} suspension had an iron stock concentration of approximately $7.8 \pm 0.1\,\SI{}{\milli\gram\per\milli\liter}$ and was supplied from an open screw-thread glass vial using a \textit{Bartels mp6 micropump} along with its driver at an injection flow rate of $\SI{1.667e-7}{\meter\cubed\per\second}$. 
The SPION reservoir was connected to the micropump, and the micropump was connected to the venous cannula, via \textit{Tygon LMT-55} tubing. 
This tubing, which had an inner diameter of $\SI{1.59}{\milli\meter}$ and an outer diameter of $\SI{3.18}{\milli\meter}$, was used throughout the entire testbed.
        
\subsubsection{Communication Channel} The communication channel begins at the \ac{TX} and ends at the center of the \ac{RX}. While only channels comprised of single pipes were considered in~\cite{Bartunik2023}, in this work, we additionally investigated branched channel topologies that provide multiple paths from the \ac{TX} to the \ac{RX}, see Figs.~\ref{fig:TestbedSetup}b)--\ref{fig:TestbedSetup}e). The branched channels were implemented using plastic y-connectors with an inner diameter of $\SI{1.59}{\milli\meter}$. To prevent inter-experiment flow rate variations caused by elevation changes in the tubing, all channel pipes were arranged on a flat plastic surface.
    
\subsubsection{Inductive Receiver} 
\begin{figure}
        \centering
        \includegraphics[width=\textwidth]{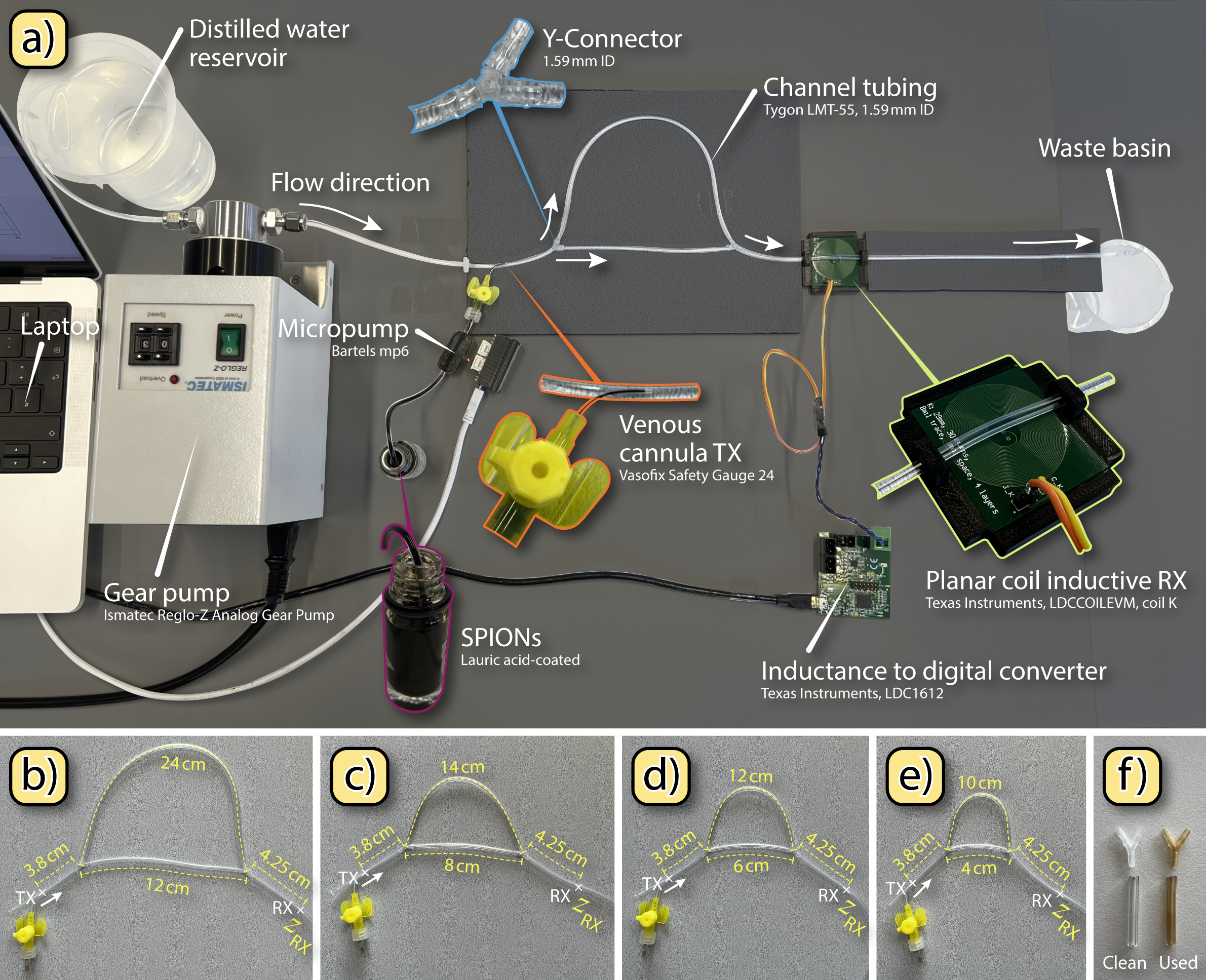}
        \caption{\textbf{Branched \ac{SPION}-testbed.} \textbf{a)} Testbed components. \textbf{b)--e)}~Branched channel topologies. \textbf{f)} Tubing material before use and after days of use, showing indications for \ac{SPION} adsorption to the tube walls.}
        \label{fig:TestbedSetup}
    \end{figure}

As a transparent inductive \ac{RX}, we employed the four-layer circular planar coil “K” from the \textit{LDCCOILEVM Reference Coil Board} by Texas Instruments, featuring an inner diameter of $\SI{8}{\milli\meter}$ and an outer diameter of $\SI{29}{\milli\meter}$. This coil was placed directly below the tubing and connected to an LC-oscillation circuit from Texas Instruments, which incorporates the \textit{LDC1612 Inductance-to-Digital Converter}, see Fig.~\ref{fig:TestbedSetup}a). The sensor output a signal in terms of resonance frequency $f_\mathrm{res}(t)$, expressed in unit $\SI{}{\hertz}$. For details on the sensing mechanism, we refer to the \ac{RX} model in Subsection~\ref{ssec:Inductive_Receiver_Model}.

\subsubsection{Waste Basin} After passing the \ac{RX}, the \ac{SPION}-distilled water mixture continued through the remaining tubing and ultimately dripped into a plastic waste basin. To ensure reproducible experimental conditions and consistent flow rates within the testbed, the basin was kept physically disconnected from the tubing.

\subsection{Measurement Procedure, Post-Processing, and Data Availability}\label{ssec:Measurement_Procedure}

\begin{figure}
        \centering
        \includegraphics[width=.75\textwidth]{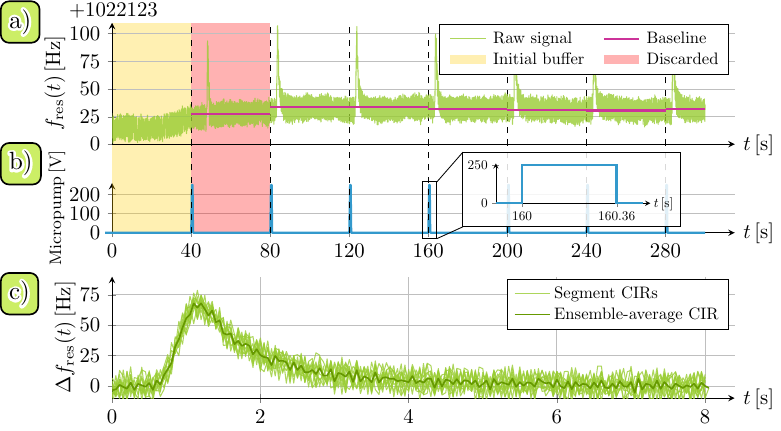}
        \caption{\textbf{Measurement procedure.} \textbf{a)} Repeated \ac{CIR} measurements. Repetitions are of $\SI{40}{\second}$ duration to avoid \ac{ISI}. Initial transient effects in the testbed are omitted through an initial buffer interval (yellow) and the discarding of the first measured \ac{CIR} (red). \textbf{b)} Micropump injections occur over a $\SI{360}{\milli\second}$ period.
        \textbf{c)} The resulting ensemble-average \ac{CIR} for a single pipe channel.}
        \label{fig:Measurement_Procedure}
\end{figure}

To measure a single experimental \ac{CIR}, we first stirred the \ac{SPION} solution in its reservoir to ensure that the suspension was as homogeneous as possible, particularly with respect to particle agglomerations that may have formed. 
We subsequently injected a rectangular \ac{SPION} pulse of duration\footnote{This duration was chosen to ensure that a sufficient number of \acp{SPION} was injected to allow for measurable responses in long and branched channels, while still approximating an impulsive release.} $T_\mathrm{inj}=\SI{360}{\milli\second}$, cf.~blue curve in~Fig.~\ref{fig:Measurement_Procedure}, i.e.,
\begin{equation}\label{eqn:InjectionFunction}
        \lambda (t) = \dfrac{N}{T_{\mathrm{inj}}}\mathrm{rect}\left( \dfrac{t-T_{\mathrm{inj}}/2}{T_{\mathrm{inj}}}\right)\,\text{,}
\end{equation}
where $N=\SI{4e12}{}$~\cite{Bartunik2023}.
The resulting sensor response was recorded over time.
    
A typical measurement series is depicted in~Fig.~\ref{fig:Measurement_Procedure}, including the raw sensor data in Fig.~\ref{fig:Measurement_Procedure}a), the micropump activation signal in Fig.~\ref{fig:Measurement_Procedure}b), and the average \ac{CIR} in Fig.~\ref{fig:Measurement_Procedure}c).
To obtain the average \ac{CIR}, by default, the injection process in~\eqref{eqn:InjectionFunction} is repeated 12 times, with $\SI{40}{\second}$ between successive activations of the micropump to prevent \ac{ISI}. 
The individual $\SI{40}{\second}$ intervals are then segmented, as indicated by the dashed vertical lines in Fig.~\ref{fig:Measurement_Procedure}a), and the corresponding sensor baseline is subtracted from each segment to yield \acp{CIR} in terms of resonance frequency \textit{shift} $\Delta f_\mathrm{res}(t)$ in unit $\SI{}{\hertz}$. 
From the last eleven\footnote{One of the eleven \ac{CIR} segments was discarded for two channel topologies presented in Section~\ref{sec:Results} due to measurement disturbances caused by air bubbles in the fluid.} individual segment \acp{CIR}, we compute an ensemble-average \ac{CIR}, cf.~Fig.~\ref{fig:Measurement_Procedure}c). The baseline for each segment is defined as the average sensor response over the time interval $[\SI{25}{\second}, \SI{30}{\second}]$ within the corresponding $\SI{40}{\second}$ segment, during which all \acp{SPION} are assumed to have been flushed from the channel, cf.~purple horizontal lines in~Fig.~\ref{fig:Measurement_Procedure}. 
This baseline correction helps counteract sensor drift, which typically occurs over the duration of several minutes or hours due to changes in the \ac{EM} environment of the testbed. 
To avoid initial transient effects, each measurement series begins with $\SI{40}{\second}$ initial dead time and the following segment containing the first \ac{SPION} injection is always discarded, cf.~yellow and red area in Fig.~\ref{fig:Measurement_Procedure}a), respectively.
The default experimental parameter values are summarized in Table~\ref{tab:Default_Parameters_Testbed}.
\begin{table}
    \centering
    \caption{Default Testbed Parameters.}
    \resizebox{.9\textwidth}{!}{%
    \begin{tabular}{|c||l|l|l|}
        \hline
        & \textbf{Parameter} & \textbf{Value} & \textbf{Description} \\
        \hline\hline
        \multirow{4}{*}{\rotatebox[origin=c]{90}{\textbf{\acp{SPION}}}}
        & $R$ & $\SI{19.2}{\nano\meter}$ & \ac{SPION} hydrodynamic radius \\ \cline{2-4}
        & $\chi_\mathrm{ref}$ & $\SI{5.99e-3}{}$ & \ac{SPION} magnetic susceptibility \\ \cline{2-4}
        &  & $\SI{7.8}{\milli\gram\per\milli\liter}$ & \ac{SPION} solution iron concentration \\ \cline{2-4}
        & $N$ & $\SI{4e12}{}$ & Number of \acp{SPION} per injection \\ 
        \hline\hline
        \multirow{6}{*}{\rotatebox[origin=c]{90}{\textbf{Tubing \& Flow}}}
        & $Q$ & $\SI{2.861e-7}{\meter\cubed\per\second}$ & Background flow rate \\ \cline{2-4}
        &  & $\SI{1.667e-7}{\meter\cubed\per\second}$ & Injection flow rate \\ \cline{2-4}
        & $r$ & $\SI{0.795}{\milli\meter}$ & Inner radius of tubing and y-connector \\ \cline{2-4}
        & & $\SI{3.18}{\milli\meter}$ & Outer diameter of tubing \\ \cline{2-4}
        & & $\SI{0.7}{\milli\meter}$ & Inner diameter of venous cannula \\
        \cline{2-4}
        & & $\SI{19}{\milli\meter}$ & Length of venous cannula \\ \hline\hline 
        \multirow{5}{*}{\rotatebox[origin=c]{90}{\textbf{Protocol}}}
        & & $\SI{40}{\second}$ & Initial buffer duration \\ \cline{2-4}
        & $T_{\mathrm{inj}}$ & $\SI{360}{\milli\second}$ & Duration of injection pulse \\ \cline{2-4}
        & & $\SI{40}{\second}$ & Time between consecutive injections \\ \cline{2-4}
        & & $[\SI{25}{\second}, \SI{30}{\second}]$ & Interval within segment for baseline calculation \\
        \cline{2-4}
        & & $11$ & Number of \ac{CIR} repetitions for averaging \\ \hline\hline 
        \multirow{6}{*}{\rotatebox[origin=c]{90}{\textbf{Sensor}}}
        & $C$& $\SI{68}{\pico\farad}$ & LC oscillation circuit capacitance\\ \cline{2-4}
        & $L_0$& $\SI{206.227}{\micro\henry}$ & Default coil inductance\\ \cline{2-4}
        & & $\SI{8}{\milli\meter}$ & Inner diameter of coil\\ \cline{2-4}
        & $d_{\mathrm{coil}}$ & $\SI{29}{\milli\meter}$ & Outer diameter of coil\\ \cline{2-4}
        & & $\SI{0}{\milli\meter}$ & Distance between coil and pipe wall\\ \cline{2-4}
        & $\sigma^2$& $\SI{63.89}{\hertz\squared}$ & Signal-independent sensor noise power\\ \hline\hline
        & $T$ & $\SI{23}{\degreeCelsius}$ & Approximate room temperature \\
        \hline
    \end{tabular}
    }
    \label{tab:Default_Parameters_Testbed}
\end{table}

In line with open science principles, all experimental measurements presented in this work are openly accessible via Zenodo under the CC BY 4.0 license at \url{https://doi.org/10.5281/zenodo.17581905}. Researchers using this dataset are kindly asked to cite it using the associated Zenodo \ac{DOI}~\cite{jakumeit2025vessel}.

\subsection{Inductive Receiver Model}\label{ssec:Inductive_Receiver_Model}
Due to their magnetic properties, \acp{SPION} can be detected using capacitive and inductive sensors. 
In this work, we model the planar coil inductive \ac{RX} used in the testbed in~\cite{Bartunik2023}. 
Unlike conventional cylindrical coils, planar coils can be positioned adjacent to the vessels, eliminating the need to envelop them, cf.~Fig.~\ref{fig:TestbedSetup}. 
When \acp{SPION} enter the vicinity of the coil, a cascade of physical processes leads to the detection of the molecules through the oscillator, as detailed in the following.

\begin{figure}
    \centering
    \includegraphics[width=.6\textwidth]{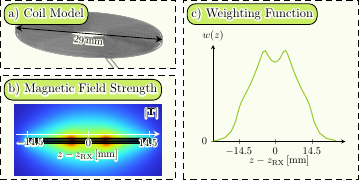}
    \caption{\textbf{Planar coil COMSOL simulation.} \textbf{a)}~\Acf{3-D} multi-layered planar coil geometry in COMSOL. \textbf{b)}~Simulated magnetic field strength $|\pmb{\Upsilon}|$ along a slice through the center of the coil. \textbf{c)}~Resulting weighting function $w(z)$ $\SI{1}{\milli\meter}$ above the coil surface within the slice shown in b).}
    \label{fig:COMSOL_Planar_Coil_Simulations}
\end{figure}

\subsubsection{Magnetic Susceptibility} In the presence of \acp{SPION}, the volume magnetic susceptibility $\chi_\mathrm{v}$ (unitless) of the environment increases, since water and air exhibit much lower susceptibilities compared to the nanoparticles.
Given that the magnetic field around planar coils is non-homogeneous, cf.~Figs.~\ref{fig:COMSOL_Planar_Coil_Simulations}b) and ~\ref{fig:COMSOL_Planar_Coil_Simulations}c), different \acp{SPION} contribute differently to the received signal, depending on their position.
In particular, their contribution is proportional to the strength of the magnetic field\footnote{For simplicity, we assume the \ac{RX} pipe diameter is small enough, such that all \acp{SPION} at a given $z$ experience the same magnetic field strength.}, denoted by $|\pmb{\Upsilon} (z)|$, which we capture using the unitless weighting function~\cite{Wicke2022}
\begin{equation}\label{eqn:CoilWeightingFunction}
	w(z)=\beta \chi_\mathrm{ref} |\pmb{\Upsilon}(z)|\,\text{,}
\end{equation}
where $\beta$ denotes a proportionality constant. 
Since closed-form expressions for the magnetic field strength can only be found for simple coil geometries, we obtain $|\pmb{\Upsilon}(z)|$ through COMSOL\textsuperscript{\textregistered} simulations\footnote{\label{foot:COMSOLSimulationFile}For the simulation file, see~\url{https://doi.org/10.5281/zenodo.13744883}.}, cf.~Fig.~\ref{fig:COMSOL_Planar_Coil_Simulations}. 
Subsequently, the volume magnetic susceptibility $\chi_\mathrm{v}(t)$ is obtained from the integral in~\eqref{eqn:ReceivedSignal}, i.e., $\chi_\mathrm{v}(t)=\int_{z\in\mathrm{dom}(w(\cdot ))}w(z)c(z,t)\,\mathrm{d}z$, using the weighting function in~\eqref{eqn:CoilWeightingFunction}.

\subsubsection{Inductance} A change in $\chi_\mathrm{v}(t)$ in turn implies a change in the coil inductance
\begin{equation}\label{eqn:MSPCInductance}
	L(t)=\mu_\mathrm{r}(t)L_0=\left(1+\chi_\mathrm{v}(t) \right)L_0\,\text{,}
\end{equation}
where $\mu_\mathrm{r}(t)$ is a unitless quantity denoting the relative permeability and $L_0$ denotes the coil inductance when no \acp{SPION} are in the proximity. 
COMSOL simulations\footnoteref{foot:COMSOLSimulationFile} yield $L_0=\SI{206.51}{\micro\henry}$, which is very close to the value $L_0=\SI{206.227}{\micro\henry}$ reported by the coil manufacturer, which we use in the following. 

\subsubsection{Resonance Frequency Shift} The LC oscillation circuit in~\cite{Bartunik2023} employs a capacitor with capacitance $C=\SI{68}{\pico\farad}$ and resonates at frequency
\begin{equation}\label{eqn:NaturalResonanceFrequency}
	f_\mathrm{res}(t)=\dfrac{1}{2\pi \sqrt{L(t)C}}\,\text{.}
\end{equation}
Let $f_{\mathrm{res},0}$ denote the resonance frequency if no \acp{SPION} are near the coil, i.e., for $L(t)=L_0$. 
Then, we define the deterministic received signal as resonance frequency shift~\cite[Eq.~(1)]{Bartunik2022}
\begin{equation}\label{eqn:ResonanceFrequencyShift}
	r(t)=\Delta f_\mathrm{res}(t)=f_{\mathrm{res},0}-f_\mathrm{res}(t)
=\dfrac{\sqrt{L(t)}-\sqrt{L_0}}{2\pi\sqrt{L_0L(t)C}}\,\text{.}
\end{equation}
Consequently, complying with the generic framework, the received signal $r(t)$ for \ac{SPION}-based \ac{MC} results from~\eqref{eqn:ReceivedSignal} with $w(z)$ in~\eqref{eqn:CoilWeightingFunction} and $f(x)=\left(\sqrt{(1+x)L_0}-\sqrt{L_0}\right)\big/\left(2\pi\sqrt{L_0^2(1+x)C}\right)$ with $x=\chi_\mathrm{v}(t)$.

\subsubsection{Noise Model}
In practice, the received signal is impacted by noise. 
To characterize the noise at the planar coil \ac{RX}, we evaluate the measured \acp{CIR}. 
For each testbed channel setting, we collect the deviations between eleven measured \acp{CIR} and their respective ensemble-averaged \ac{CIR} over time, and analyze the distribution of the deviations.
We observe similar signal-independent, additive white Gaussian noise $n_\mathrm{sensor}$ across all settings and times, which is accurately modeled~as
\begin{equation}\label{eqn:Gaussian_Noise}
	n_\mathrm{sensor}\sim \mathcal{N}(\xi =\SI{0}{\hertz},\sigma^2=\SI{63.89}{\hertz\squared})\,\text{,}
\end{equation}
where $\mathcal{N} (\xi,\sigma^2)$ denotes the Gaussian distribution with mean $\xi$ and variance $\sigma^2$.
This noise, attributed to the stochastic behavior of the electrical measurement circuit, is the most dominant noise affecting the data, leading to the stochastic received signal
\begin{equation}\label{eqn:StatisticalReceivedSignal}
		\Delta F_\mathrm{res}(t)=\Delta f_\mathrm{res} (t) + n_\mathrm{sensor}(t)\,\text{,}
\end{equation}
i.e., for any $t$, $\Delta F_\mathrm{res}(t)$ is a random variable. 
Similar to~\cite[Eq.~(5)]{Jamali2017}, we define a time-average \ac{SNR}. 
In particular, for the averaging, we capture the received signal in the time interval $[t_0,t_1]$ with $t_1>t_0$, i.e.,
\begin{equation}\label{eqn:SNR}
	\begin{aligned}
	\mathrm{SNR}&=10\log_{10}\left( \dfrac{\frac{1}{t_1-t_0}\int_{t_0}^{t_1} \Delta f_\mathrm{res}^2(t)\,\mathrm{d}t}{\sigma^2}\right)\,\text{.}
	\end{aligned}
\end{equation}
The interval $[t_0, t_1]$ is appropriately chosen to capture (virtually) the entire energy of $\Delta f_\mathrm{res}(t)$.

\section{Numerical and Experimental Results}\label{sec:Results}
In this section, we first study the significance of sorption-dynamics for \ac{SPION} transport, as compared to pure advection-diffusion dynamics. Second, we present the experimental validation of the proposed end-to-end model across varying channel topologies. Third, we experimentally validate the concept of the dispersion space. Lastly, we illustrate the usefulness of the dispersion space for the case of more complex synthetic \acp{VN}.

\subsection{Transport With and Without Sorption Dynamics}
\begin{figure*}
    \centering
    \includegraphics[width=\textwidth]{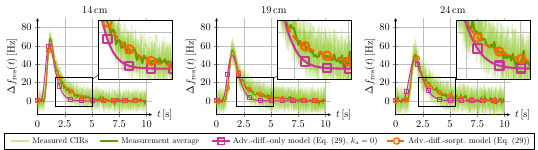}
    \caption{\textbf{Advection-diffusion-only vs.~advective-diffusive-sorptive end-to-end model.} Model and experimental \acp{CIR} for three exemplary straight channel topologies. To show the best possible fit of each model for each scenario, each model curve was individually fitted to the respective measured average curve. For the $\{\SI{14}{},\SI{19}{},\SI{24}{}\}\,\SI{}{\centi\meter}$ settings, the fitted parameters of the advection-diffusion-only model are $\alpha=\{\SI{1.197}{},\SI{1.247}{},\SI{1.081}{}\}\times 10^{-3}$ and $\beta=\{\SI{3.265}{},\SI{3.061}{},\SI{3.210}{}\}\times 10^{-18}$, respectively. The corresponding fitted parameters of the advective-diffusive-sorptive model are $\alpha=\{\SI{4.675}{},\SI{3.163}{},\SI{3.135}{}\}\times 10^{-3}$, $\beta=\{\SI{2.872}{},\SI{2.859}{},\SI{3.016}{}\}\times 10^{-18}$, $k_\mathrm{a}=\{\SI{0.637}{},\SI{0.341}{},\SI{0.337}{}\}\,\SI{}{\per\second}$, and $k_\mathrm{d}=\{\SI{1.372}{},\SI{0.789}{},\SI{0.858}{}\}\,\SI{}{\per\second}$, respectively.}
    \label{fig:Advective_Diffusive_vs_Sorptive_Transport}
\end{figure*}

Sorption dynamics significantly influence \ac{SPION} transport in \acp{VN}, both in our synthetic setup using silicon tubing, and likely, under different conditions, in biological vasculature.
Experimentally, their relevance is evident from the prolonged clearance times observed after impulsive \ac{SPION} injections, as well as the persistent brown discoloration of tubing material, caused by \ac{SPION} accumulation on channel walls following repeated experiments, see Fig.~\ref{fig:TestbedSetup}f).
To highlight the impact of these dynamics, in Fig.~\ref{fig:Advective_Diffusive_vs_Sorptive_Transport}, we compare experimental measurements to two end-to-end model variants: one based solely on advection and diffusion, and the proposed model that additionally accounts for sorption at channel walls. 
This comparison is performed for three representative straight channel topologies with lengths $\{\SI{14}{},\SI{19}{},\SI{24}{}\}\,\SI{}{\centi\meter}$.

For each channel topology, both the advection-diffusion-only model and the full advective-diffusive-sorptive model are individually fitted to the corresponding measured average signal to obtain the best possible fit each model can achieve in each scenario. The fitting is performed using SciPy's \textit{differential evolution} algorithm (SciPy version 1.10.1) by optimizing the eddy diffusion coefficient via $\alpha$, the scale parameter $\beta$ of the magnetic field weighting function $w(z)$, and the sorption rate constants $k_\mathrm{a}$ and $k_\mathrm{d}$. The fitted parameter values are listed in the caption of Fig.~\ref{fig:Advective_Diffusive_vs_Sorptive_Transport}.

The results show that, across all channel lengths, the advection-diffusion-only model cannot reproduce the heavy signal tails observed in the measurements, even under optimal fitting. In contrast, the proposed model including sorption closely matches the experimental data, accurately reflecting the influence of molecule-wall interactions and confirming the critical role of sorption in \ac{SPION} transport.

\subsection{Experimental Model Validation}\label{ssec:Experimental_Model_Validation}

To validate the end-to-end model introduced in Section~\ref{sec:System_Model} and Subsection~\ref{ssec:Inductive_Receiver_Model}, we compare the predicted \acp{CIR} from~\eqref{eqn:ResonanceFrequencyShift} to experimentally measured \acp{CIR} across nine distinct channel topologies. Specifically, we consider single-vessel channels of lengths $\{ 9,14,19,24,29\}\,\SI{}{\centi\meter}$ as well as the branched configurations shown in Figs.~\ref{fig:TestbedSetup}b)–e). All measurements were conducted as described in Subsection~\ref{ssec:Measurement_Procedure}, and all model predictions are based on the default testbed parameters listed in Table~\ref{tab:Default_Parameters_Testbed}, along with the injection function defined in~\eqref{eqn:InjectionFunction}.
The model includes four parameters that must be estimated from experimental data, as directly measuring the parameters is infeasible: the sorption rate constants $k_\mathrm{a}$ and $k_\mathrm{d}$, the eddy-diffusion coefficient proportionality factor $\alpha$, and the electromagnetic scaling factor $\beta$, which accounts for deterministic but unknown environmental influences on the inductive coil \ac{RX}. For parameter fitting, we use the average measured \ac{CIR} from the branched channel in Fig.~\ref{fig:TestbedSetup}d) and employ SciPy's differential evolution algorithm, yielding a single unified parameter set $k_\mathrm{a} = \SI{0.3699}{\per\second}$, $k_\mathrm{d} = \SI{0.9362}{\per\second}$, $\alpha = 0.008289$, and $\beta = \SI{2.441e-18}{}$, which is subsequently used for all model predictions across all topologies.
\begin{figure*}
    \centering
    \includegraphics[width=\linewidth]{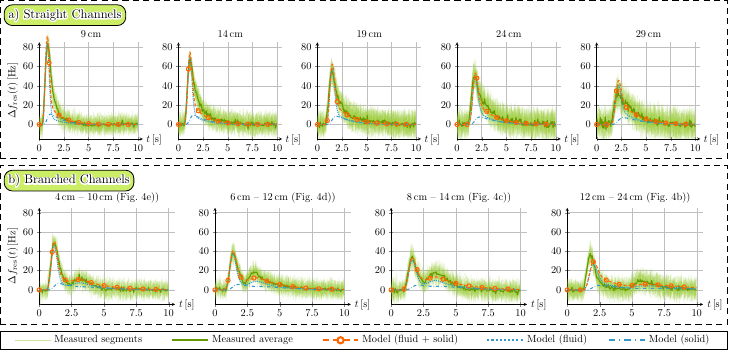}
    \caption{\textbf{Comparison of \ac{SPION} measurements and end-to-end model predictions across nine channel topologies.} Results for straight and branched channels are shown in a) and b), respectively. For all scenarios, fluid phase, solid phase, and combined phase (fluid + solid) model predictions are shown. Parameters $k_\mathrm{a} = \SI{0.3699}{\per\second}$, $k_\mathrm{d} = \SI{0.9362}{\per\second}$, $\alpha = 0.008289$, and $\beta = \SI{2.441e-18}{}$ were jointly fitted to the data.}
    \label{fig:Experimental_Model_Validation}
\end{figure*}
Fig.~\ref{fig:Experimental_Model_Validation} summarizes the validation results. For each channel topology, individual measured realizations and the average measured \ac{CIR} are shown in light and dark green, respectively. The model-predicted contributions of signaling molecules in the fluid and solid phases are shown as blue dotted and dash-dotted curves, respectively. The total predicted received signal, combining both phases, is depicted by the orange dashed curve.

Overall, the model shows good agreement with the experimental data. For both straight channels of varying lengths (Fig.~\ref{fig:Experimental_Model_Validation}a) and branched topologies (Fig.~\ref{fig:Experimental_Model_Validation}b), peak times and amplitudes are generally well reproduced. 
In most cases, the heavy tails observed in measurements, caused by sorption at channel walls, are also accurately captured. 
Note that some mismatch between model predictions and experiments is inevitable, as \acp{SPION} tend to form particle clusters through agglomeration, which alters diffusion and sorption dynamics in unpredictable ways\footnote{Close-to-perfect agreement between model and experiment can be achieved by individually fitting the model parameters $k_\mathrm{a}$, $k_\mathrm{b}$, $\alpha$, and $\beta$ to measurements from \textit{individual} channel topologies, as opposed to using a unified parameter set, cf., e.g., Fig.~\ref{fig:Advective_Diffusive_vs_Sorptive_Transport}.}.
The model further reveals that \acp{SPION} in the fluid phase contribute more strongly to the received signal than those bound to the wall (solid phase).
Individual \ac{CIR} measurements exhibit considerable noise, primarily due to dominant sensor noise. The assumption of signal-independent noise, as used in~\eqref{eqn:StatisticalReceivedSignal}, is visually supported by the approximately constant noise power across time within each setting. This consistency is expected, as each measurement series for a given setting lasts only a few minutes. 
In contrast, the time between different measurement series can span several hours, during which environmental conditions may change. 
Consequently, noise levels may vary across topologies (e.g., between the $\SI{9}{\centi\meter}$ and $\SI{29}{\centi\meter}$ straight channels). 
Given the high sensitivity of the inductive coil \ac{RX}, even small or distant environmental changes can noticeably affect the noise power.
In summary, the results demonstrate the strong predictive power of the model, as it accurately reproduces measured signals across diverse straight and branched channel topologies using a single parameter set obtained from just one calibration measurement.

\subsection{Experimental Dispersion Space Validation}
\begin{figure}
    \centering
    \includegraphics[width=.75\textwidth]{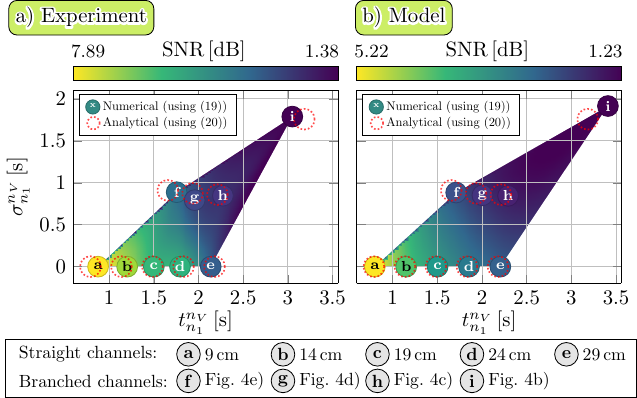}
    \caption{\textbf{Experimental and model dispersion space.} Dispersion space for all nine channel topologies from Fig.~\ref{fig:Experimental_Model_Validation}, using \textbf{a)} experimental data and \textbf{b)} model predictions. Each \ac{VN} is represented by two markers: a colored marker ($t_{n_1}^{n_V}$ and $\sigma_{n_1}^{n_V}$ evaluated using the numerically determined peak times in~\eqref{eqn:PipeTmax}) and a red dotted marker ($t_{n_1}^{n_V}$ and $\sigma_{n_1}^{n_V}$ evaluated using the approximate analytical peak time in~\eqref{eqn:PipeTmaxApproximation}). The marker color indicates the corresponding \ac{SNR}, highlighting a strong correlation between increased dispersion, i.e., increased $t_{n_1}^{n_V}$ and $\sigma_{n_1}^{n_V}$, and lower \ac{SNR}. The letters in the numerical markers correspond to the channel topologies listed below the plots.}
    \label{fig:Dispersion_Space_Results}
\end{figure}

To verify that the received \ac{SNR} can be inferred directly from the \ac{VN} topology using the dispersion metrics introduced in Section~\ref{sec:Topology_Dispersion_Metrics}, we locate all nine channel topologies from Fig.~\ref{fig:Experimental_Model_Validation} in the dispersion space using the metrics defined in~\eqref{eqn:NetworkAverageTimeOfArrivalWeightedAverage} and~\eqref{eqn:PTV}, and evaluate their \acp{SNR} based on measurements and model predictions, respectively. We refer to these as the \textit{experimental} and \textit{model} dispersion spaces, respectively. For each \ac{VN}, in each dispersion space, two markers are plotted at the positions obtained from the metrics in~\eqref{eqn:NetworkAverageTimeOfArrivalWeightedAverage} and~\eqref{eqn:PTV}, once evaluated analytically and once numerically, see caption of Fig.~\ref{fig:Dispersion_Space_Results}.
Marker colors indicate the \ac{SNR} of the received signal in the corresponding \ac{VN}.
Note that we interpolate the \ac{SNR} between the colored markers for better visual representation of the relationship between the location of a \ac{VN} in the dispersion space and its \ac{SNR}.

The \ac{SNR} in the \textit{experimental} dispersion space is computed by considering the deviation of individual measured \ac{CIR} realizations from the average measured signal as noise. The average noise power $\sigma^2 = \SI{63.89}{\hertz\squared}$ is determined across all nine topologies. For each topology, the \textit{experimental} \ac{SNR} is then evaluated using~\eqref{eqn:SNR}, where $\Delta f_\mathrm{res}(t)$ is the average measured \ac{CIR} and $\sigma^2 = \SI{63.89}{\hertz\squared}$, cf.~\eqref{eqn:Gaussian_Noise}. Since all signals in Fig.~\ref{fig:Experimental_Model_Validation} cover roughly the same time interval, we use the fixed interval $[t_0=\SI{0}{\second},t_1=\SI{10}{\second}]$ to compute the \ac{SNR} in~\eqref{eqn:SNR}.
As discussed in Subsection~\ref{ssec:Experimental_Model_Validation}, the sensor noise level can vary between measurements due to environmental changes. Averaging the noise power across all settings mitigates these effects, allowing the resulting \ac{SNR} to reflect primarily the influence of the underlying topology.
The \ac{SNR} in the \textit{model} dispersion space is similarly obtained via~\eqref{eqn:SNR}, using the model-predicted $\Delta f_\mathrm{res}(t)$ and the same uniform noise power $\sigma^2 = \SI{63.89}{\hertz\squared}$ for consistency across all scenarios.

Experimental and model results are shown in Figs.~\ref{fig:Dispersion_Space_Results}a) and~\ref{fig:Dispersion_Space_Results}b), respectively. We observe that, both in the experimental and model dispersion space, a strong correlation and clear pattern between
the position of each \ac{VN} in the space and its corresponding \ac{SNR} emerges.
Specifically, the \ac{SNR} decreases systematically with increasing $t_{n_1}^{n_V}$ and $\sigma_{n_1}^{n_V}$, validating the effectiveness of the proposed metrics in capturing dispersion-induced signal degradation. The extremes are intuitive: The shortest straight channel (marker “a”) yields the highest \ac{SNR}, while the largest branched network (marker “i”) exhibits the lowest \ac{SNR}, consistent across both experimental and model results. Note, also, that only the branched topologies exhibit a non-zero $\sigma_{n_1}^{n_V}$, since multi-path propagation does not occur in straight channels.

Additionally, the numerically evaluated markers (colored) generally align well with their corresponding analytical counterparts (red dashed), suggesting that for the fitted sorption rates ($k_\mathrm{a} = \SI{0.3699}{\per\second}$, $k_\mathrm{d} = \SI{0.9362}{\per\second}$), the position in the dispersion space is predominantly governed by advection and diffusion. 
Only for the largest branched channel topology, represented by marker "i", a significant mismatch in the position of the colored and dashed markers is observed. 
In both the experimental and model dispersion space, this happens because the effects of the sorption dynamics become more pronounced as travel times increase, i.e., for larger topologies, and the advection-diffusion-only model does not capture these dynamics.
Moreover, in the experimental space, slight lateral deviations between numerical and analytical markers arise from small mismatches in peak times (typically within $\SI{100}{\milli\second}$). 
In contrast, in the model space, numerical markers only shift rightward, as they incorporate slight signal delays caused by sorption.

Comparing the experimental and model \acp{SNR}, we observe that the model \acp{SNR} are generally slightly lower than the experimental \acp{SNR}. 
As seen in Fig.~\ref{fig:Experimental_Model_Validation}, this can be attributed to the model \acp{CIR} partially lying below the measured average curves, while both model and experiment assume the same noise power.
Nonetheless, measured \acp{SNR} ($1.38\,\mathrm{dB}-7.89\,\mathrm{dB}$) and model \acp{SNR} ($1.23\,\mathrm{dB}-5.22\,\mathrm{dB}$) are overall comparable. 
In summary, the results suggest that the proposed dispersion space can serve as a useful tool for estimating the expected signal quality based on the \ac{VN} topology, both analytically and experimentally.
In particular, the proposed metrics in~\eqref{eqn:NetworkAverageTimeOfArrivalWeightedAverage} and~\eqref{eqn:PTV} not only capture the topology-\ac{SNR} link for the extreme cases (markers "a" and "i"), but also accurately predict the \ac{SNR} trend of all other intermediate \acp{VN}.

\subsection{Model Dispersion Space for Complex Vessel Networks}

\begin{figure}
    \centering
    \includegraphics[width=0.65\linewidth]{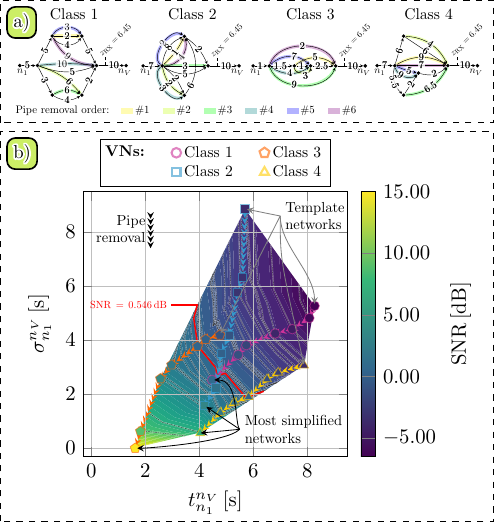}
    \caption{\textbf{Model dispersion space with 28 complex \acp{VN}.}
\textbf{a)} Four template \acp{VN} (classes~1-4) are shown. For all \acp{VN}, $l_i$ are given in $\SI{}{\centi\meter}$, $r_i = \SI{1}{\milli\meter}$, and $z_\mathrm{RX} = \SI{6.45}{\milli\meter}$. From the templates, 28 \acp{VN} (including the templates) are generated by successively removing pipes in the order indicated by the colored edges.
\textbf{b)} All \acp{VN} from a) are mapped into the dispersion space, with the received \ac{SNR} color-coded. The impact of pipe removal is visualized via arrows. Template \acp{VN} and their most simplified counterparts are highlighted. Between markers, \ac{SNR} values are interpolated, and dashed black lines depict \ac{SNR} contours, one per marker. The red contour line indicates an exemplary \ac{SNR} threshold (e.g., as potentially required by a sensor): \acp{VN} to the lower left satisfy the threshold, those to the upper right do not. All simulations assume $\alpha=\SI{8.289e-03}{}$, $\beta=\SI{2.441e-18}{}$, $k_\mathrm{a} = \SI{0}{\per\second}$, and $\lambda(t) = N\delta(t)$; default parameters apply otherwise.}
    \label{fig:dispersion_space_with_complex_vessel_networks}
\end{figure}

To further explore the usefulness of the dispersion space beyond the \acp{VN} considered so far, in Fig.~\ref{fig:dispersion_space_with_complex_vessel_networks}, more complex \acp{VN}, as compared to Fig.~\ref{fig:Dispersion_Space_Results}, are simulated using~\eqref{eqn:ResonanceFrequencyShift}. 
To this end, we consider four different classes of \acp{VN}, as illustrated in Fig.~\ref{fig:dispersion_space_with_complex_vessel_networks}a).
For each of the four template \acp{VN} (classes 1--4) shown in Fig.~\ref{fig:dispersion_space_with_complex_vessel_networks}a), six
further networks are generated by iteratively removing one
pipe in the order indicated in color, totaling 28 \acp{VN}.
The pipe lengths in unit $\SI{}{\centi\meter}$ are annotated to the corresponding edges in Fig.~\ref{fig:dispersion_space_with_complex_vessel_networks}a), with a radius of $r_i=\SI{1}{\milli\meter}$ and the \ac{RX} coil centered at $z_\mathrm{RX}=\SI{6.45}{\centi\meter}$ in the outlet pipe.
In each \ac{VN}, we evaluate the proposed model for the
instantaneous injection of $N=\SI{4e12}{}$ \acp{SPION}, i.e., $\lambda (t)=N\delta(t)$, using $\alpha=\SI{8.289e-03}{}$, $\beta=\SI{2.441e-18}{}$, and $k_\mathrm{a} = \SI{0}{\per\second}$. In Fig.~\ref{fig:dispersion_space_with_complex_vessel_networks}b), all 28 \acp{VN} are mapped into the dispersion space as markers by evaluating~\eqref{eqn:NetworkAverageTimeOfArrivalWeightedAverage} and~\eqref{eqn:PTV} using the peak time expression in~\eqref{eqn:PipeTmaxApproximation}, which is accurate for $k_\mathrm{a} = \SI{0}{\per\second}$. 
Additionally, the received \ac{SNR} with $\sigma^2=\SI{63.89}{\hertz\squared}$ for each \ac{VN} is computed using~\eqref{eqn:SNR}, where $t_0$ and $t_1$ correspond to the times at which $\Delta f_\mathrm{res}(t)$ first exceeds and last falls below the small value of $\SI{0.1}{\hertz}$, respectively. 
This adaptive approach to defining the \ac{SNR} interval ensures that, even when signal energy is distributed across widely differing time regions, the selected interval consistently encompasses all relevant energy contributions. This is particularly important for the diverse \acp{VN} in Fig.~\ref{fig:dispersion_space_with_complex_vessel_networks}a), whose varying topologies yield signals that fall into vastly distinct temporal ranges. Calculated \acp{SNR} are then color-coded, and for each \ac{VN}, an \ac{SNR} contour line is plotted.

As in Fig.~\ref{fig:Dispersion_Space_Results}, Fig.~\ref{fig:dispersion_space_with_complex_vessel_networks}b) reveals a clear pattern between the location of a \ac{VN} in the dispersion space and its corresponding \ac{SNR}. The contour lines run largely parallel, indicating a consistent \ac{SNR} gradient across the space. As pipes are sequentially removed, \acp{VN} shift closer to the origin, reflecting reduced dispersion and increased \ac{SNR}. Since each pipe in a given topology generally transports a different fraction of the total number of injected molecules, and thus has a different impact on the total received signal, the magnitude of the positional shifts in the dispersion upon pipe removal may differ between individual removals.
Notably, different \ac{VN} classes cluster in distinct regions of the dispersion space, and the simplified variants of each class remain topologically coherent. Despite the class-specific differences, the proposed metrics effectively capture the fundamental relationship between topology and \ac{SNR}.

Given a sensor with a specific \ac{SNR} requirement, the dispersion space provides a means to predict which \acp{VN} can precede the sensor while still satisfying this constraint. 
In Fig.~\ref{fig:dispersion_space_with_complex_vessel_networks}b), a red contour line illustrates an exemplary threshold of $\mathrm{SNR} = \SI{0.546}{\deci\bel}$. 
If the \ac{VN} upstream of the sensor is located to the lower left of this line, the \ac{SNR} requirement is met, enabling reliable sensing. Conversely, \acp{VN} located to the upper right fall below the required \ac{SNR}, indicating that sensing would not be successful for those configurations.
In in-body \ac{CVS} applications, the upstream environment of the sensor depends on its anatomical location, resulting in varying \acp{VN} observed by the sensor. The proposed framework provides a principled way to evaluate whether a given placement can meet given \ac{SNR} requirements. Similarly, when designing a fluidic \ac{MC} testbed for a sensor with known sensitivity constraints, the framework can inform which channel topologies are viable before physical implementation, ensuring the sensor operates within its performance limits.

\section{Conclusion}\label{sec:Conclusion}
In this paper, we presented a novel channel model for advection-diffusion-sorption-governed \ac{MC} in \acp{VN}. 
Building on this model, we introduced the metrics \textit{molecule delay} and \textit{multi-path spread}, which jointly define the \textit{dispersion space} and link a given \ac{VN} topology to the degree of induced dispersion. 
This, in turn, establishes a direct link between the \ac{VN} structure and the \ac{SNR} of the molecular signal at the \ac{RX}.
We specialized the model to \ac{SPION}-based \ac{MC} by incorporating an analytical model of an inductive planar coil \ac{RX}, enabling comparisons between theoretical predictions and measurements from a branched \ac{SPION} testbed with multiple transport paths between \ac{TX} and \ac{RX}. 
This setup allowed thorough end-to-end validation of the model and confirmed the practical relevance of the dispersion space. 

Our results demonstrate that the proposed \ac{1-D} physical model, capturing advective, diffusive, and sorptive transport, can accurately reproduce experimental observations in both single vessels and \acp{VN}. Furthermore, both theory and experiment support the dispersion space as a valuable tool for estimating received signal quality based solely on \ac{VN} topology. This enables the evaluation and optimization of sensor placement strategies in complex \ac{VN} structures such as the \ac{CVS}, under specific \ac{SNR} requirements. It can also inform the design of future \ac{IoBNT} gateway devices by linking signal quality to structural features of the \ac{VN}. Additionally, the framework can guide the construction of experimental testbeds by predicting the maximum structural complexity that still ensures sufficient signal strength at the \ac{RX}.

Looking ahead, we aim to develop progressively abstracted fading models to support a more generalized understanding of \ac{MC} in \acp{VN}. Ultimately, the \ac{MC} behavior of entire organs or tissue regions may be captured using high-level statistical models. Toward this goal, future work may compare model predictions with \textit{ex vivo} or \textit{in vivo} measurements in animal models and apply the proposed framework to \acp{VN} extracted from human vasculature scans.

\appendix[Derivation of Pipe Flux]\label{sec:AppA}
Given the concentration in~\eqref{eqn:FluidPhaseConcentration}, the corresponding molecule flux for a single injected molecule can be derived as~\cite[Eq.~(4.4)]{Berg1993}
    \begin{equation}\label{eqn:PipeFluxDerivation1}
        J_i(z,t) = \dfrac{1}{N}\left(-D_i^\mathrm{eff}\dfrac{\partial c_i(z,t)}{\partial z} + \overline{u}_i c_i(z,t)\right)\,\text{.}
    \end{equation}
The derivative in~\eqref{eqn:PipeFluxDerivation1} can be expressed as
\begin{equation}\label{eqn:PipeFluxDerivation2}
    \dfrac{\partial c_i(z,t)}{\partial z} = \dfrac{\partial \tilde{c}_i(z,t)}{\partial t}\exp \left( -k_\mathrm{a} t\right) + \int\limits_0^t \dfrac{\partial \tilde{c}_i(z,t)}{\partial t} g_\mathrm{ff}(\tau, t)\,\mathrm{d}\tau\,\text{.}
\end{equation}
Moreover, it holds that~\cite[Eq.~(6)]{Jakumeit2025}
\begin{equation}\label{eqn:PipeFluxDerivation3}
    \dfrac{\partial \tilde{c}_i(z,t)}{\partial t} = \left(\dfrac{(z-\overline{u}_i t)^2}{2t} + \overline{u}_i  \right) \tilde{c}_i(z,t)\,\text{.}
\end{equation}
By inserting~\eqref{eqn:PipeFluxDerivation3} into \eqref{eqn:PipeFluxDerivation2}, and \eqref{eqn:PipeFluxDerivation2} into \eqref{eqn:PipeFluxDerivation1}, we obtain the expression in \eqref{eqn:PipeFlux}.

\section*{Acknowledgments}
We thank Prof. Dr. med. Christoph Alexiou, head of the Section of Experimental Oncology and Nanomedicine (SEON), Universitätsklinikum Erlangen, and Prof. Dr. rer. nat. Dr. habil. med. Stefan Lyer for synthesizing and providing the \acp{SPION} used in the presented experiments.

\renewcommand{\baselinestretch}{.92}
\bibliographystyle{IEEEtran}
\bibliography{references}

@Article{Mosayebi2019,
  author    = {Mosayebi, Reza and others},
  journal   = {IEEE Trans. Nanobiosci.},
  title     = {Early cancer detection in blood vessels using mobile nanosensors},
  year      = {2019},
  issn      = {1558-2639},
  month     = apr,
  pages     = {103--116},
  volume    = {18},
  doi       = {10.1109/tnb.2018.2885463},
  publisher = {Institute of Electrical and Electronics Engineers (IEEE)},
}

@Article{Simo2024,
  author    = {Simo, Cristina and others},
  journal   = {Nat. Nanotechnol.},
  title     = {Urease-powered nanobots for radionuclide bladder cancer therapy},
  year      = {2024},
  issn      = {1748-3395},
  month     = jan,
  pages     = {554--564},
  volume    = {19},
  doi       = {10.1038/s41565-023-01577-y},
  publisher = {Springer Science and Business Media LLC},
}

@Article{Akyildiz2015,
  author    = {Akyildiz, I. and Pierobon, M. and Balasubramaniam, S. and Koucheryavy, Y.},
  journal   = {IEEE Commun. Mag.},
  title     = {The {I}nternet of {B}io-{N}ano {T}hings},
  year      = {2015},
  issn      = {0163-6804},
  month     = mar,
  pages     = {32--40},
  volume    = {53},
  doi       = {10.1109/mcom.2015.7060516},
  publisher = {Institute of Electrical and Electronics Engineers (IEEE)},
}

@Article{Chahibi2013,
  author    = {Chahibi, Youssef and Pierobon, Massimiliano and Song, Sang Ok and Akyildiz, Ian F.},
  journal   = {IEEE Trans. Biomed. Eng.},
  title     = {A molecular communication system model for particulate drug delivery systems},
  year      = {2013},
  issn      = {1558-2531},
  month     = dec,
  pages     = {3468--3483},
  volume    = {60},
  doi       = {10.1109/tbme.2013.2271503},
  publisher = {Institute of Electrical and Electronics Engineers (IEEE)},
}

@Article{Bartunik2023,
  author    = {Bartunik, Max and Teller, Janina and Fischer, Georg and Kirchner, Jens},
  journal   = {IEEE Trans. Mol. Biol. Multi-Scale Commun.},
  title     = {Channel parameter studies of a molecular communication testbed with biocompatible information carriers: methods and data},
  year      = {2023},
  issn      = {2332-7804},
  month     = dec,
  number    = {4},
  pages     = {489--498},
  volume    = {9},
  doi       = {10.1109/tmbmc.2023.3325405},
  publisher = {Institute of Electrical and Electronics Engineers (IEEE)},
}

@Book{Nieuwstadt2016,
  author    = {Nieuwstadt, F. T. M.},
  editor    = {Bendiks J. Boersma and Jerry Westerweel},
  publisher = {Springer},
  title     = {Turbulence},
  year      = {2016},
  isbn      = {9783319315997},
  ppn_gvk   = {1658470842},
  subtitle  = {Introduction to Theory and Applications of Turbulent Flows},
}

@Article{Wicke2022,
  author    = {Wicke, Wayan and others},
  journal   = {IEEE Trans. Mol. Biol. Multi-Scale Commun.},
  title     = {Experimental system for molecular communication in pipe flow with magnetic nanoparticles},
  year      = {2022},
  issn      = {2332-7804},
  month     = jun,
  number    = {2},
  pages     = {56--71},
  volume    = {8},
  doi       = {10.1109/tmbmc.2021.3099399},
  publisher = {Institute of Electrical and Electronics Engineers (IEEE)},
}

@InProceedings{Bartunik2022,
  author     = {Bartunik, Max and Faghih-Naini, Samira and Maiwald, Timo and Kirchner, Jens},
  booktitle  = {Proc. ACM Int. Conf. Nanoscale Comput. Commun.},
  title      = {Planar coils for detection of magnetic nanoparticles in a testbed for molecular communication},
  year       = {2022},
  month      = oct,
  pages      = {1--6},
  collection = {NANOCOM ’22},
  doi        = {10.1145/3558583.3558844},
}

@Article{Jamali2017,
  author    = {Jamali, Vahid and Ahmadzadeh, Arman and Schober, Robert},
  journal   = {IEEE Commun. Lett.},
  title     = {On the design of matched filters for molecule counting receivers},
  year      = {2017},
  issn      = {1089-7798},
  month     = aug,
  pages     = {1711--1714},
  volume    = {21},
  doi       = {10.1109/lcomm.2017.2702178},
  publisher = {Institute of Electrical and Electronics Engineers (IEEE)},
}

@Article{Pal2023,
  author    = {Pal, Saswati and Misra, Sudip and Islam, Nabiul},
  journal   = {IEEE Trans. Mol. Biol. Multi-Scale Commun.},
  title     = {{m}-{MSC}: molecular communication-based analysis for controlled {MSC} treatment of cytokine storm},
  year      = {2023},
  issn      = {2332-7804},
  month     = sep,
  number    = {3},
  pages     = {286--294},
  volume    = {9},
  doi       = {10.1109/tmbmc.2023.3296430},
  publisher = {Institute of Electrical and Electronics Engineers (IEEE)},
}

@Article{Palanisamy2019,
  author    = {Palanisamy, Sathyadevi and Wang, Yun-Ming},
  journal   = {Dalton Trans.},
  title     = {Superparamagnetic iron oxide nanoparticulate system: {s}ynthesis, targeting, drug delivery and therapy in cancer},
  year      = {2019},
  issn      = {1477-9234},
  month     = jun,
  pages     = {9490--9515},
  volume    = {48},
  doi       = {10.1039/c9dt00459a},
  publisher = {Royal Society of Chemistry (RSC)},
}

@Article{Yu2024,
  author    = {Yu, Yafeng and others},
  journal   = {Nat. Commun.},
  title     = {Vascular network-inspired fluidic system ({VasFluidics}) with spatially functionalizable membranous walls},
  year      = {2024},
  issn      = {2041-1723},
  month     = feb,
  volume    = {15 (1437)},
  doi       = {10.1038/s41467-024-45781-3},
  publisher = {Springer Science and Business Media LLC},
}

@Book{Rappaport2024,
  author    = {Rappaport, Theodore S.},
  publisher = {Cambridge University Press},
  title     = {Wireless {C}ommunications: {P}rinciples and {P}ractice},
  year      = {2024},
  isbn      = {9781009489836},
  month     = feb,
  doi       = {10.1017/9781009489843},
}

@Book{Berg1993,
  author    = {Berg, Howard C.},
  publisher = {Princeton University Press},
  title     = {Random {W}alks in {B}iology},
  year      = {1993},
  address   = {Princeton, N.J},
  edition   = {Expanded},
  isbn      = {1400820022},
  pagetotal = {1152},
  ppn_gvk   = {1889457833},
}

@Article{Kooten1996,
  author    = {van Kooten, J.J.A.},
  journal   = {Adv. Water Resour.},
  title     = {A method to solve the advection-dispersion equation with a kinetic adsorption isotherm},
  year      = {1996},
  issn      = {0309-1708},
  month     = aug,
  number    = {4},
  pages     = {193--206},
  volume    = {19},
  doi       = {10.1016/0309-1708(95)00045-3},
  publisher = {Elsevier BV},
}

@Article{Tjabben2024,
  author  = {Tjabben, Annika and Bergkemper, Lea and Herbst, Jan and Rueb, Matthias and Lipps, Christoph and Schotten, Hans Dieter},
  journal = {Proc. IEEE Eur. Wirel. Conf.},
  title   = {Multipath signal prediction for in-body nanocommunication with volatile particles},
  year    = {2025},
  month   = mar,
  pages   = {41--46},
}

@Article{Reichold2009,
  author    = {Reichold, Johannes and Stampanoni, Marco and Keller, Anna Lena and Buck, Alfred and Jenny, Patrick and Weber, Bruno},
  journal   = {J. Cereb. Blood Flow Metab.},
  title     = {Vascular graph model to simulate the cerebral blood flow in realistic vascular networks},
  year      = {2009},
  issn      = {1559-7016},
  month     = may,
  number    = {8},
  pages     = {1429--1443},
  volume    = {29},
  doi       = {10.1038/jcbfm.2009.58},
  publisher = {SAGE Publications},
}

@Article{d’Esposito2018,
  author    = {d’Esposito, Angela and others},
  journal   = {Nat. Biomed. Eng.},
  title     = {Computational fluid dynamics with imaging of cleared tissue and of in vivo perfusion predicts drug uptake and treatment responses in tumours},
  year      = {2018},
  issn      = {2157-846X},
  month     = oct,
  number    = {10},
  pages     = {773--787},
  volume    = {2},
  doi       = {10.1038/s41551-018-0306-y},
  publisher = {Springer Science and Business Media LLC},
}

@Article{Bartunik2023a,
  author    = {Bartunik, Max and Fischer, Georg and Kirchner, Jens},
  journal   = {IEEE Trans. Mol. Biol. Multi-Scale Commun.},
  title     = {The development of a biocompatible testbed for molecular communication with magnetic nanoparticles},
  year      = {2023},
  issn      = {2332-7804},
  month     = jun,
  number    = {2},
  pages     = {179--190},
  volume    = {9},
  doi       = {10.1109/tmbmc.2023.3265565},
  publisher = {Institute of Electrical and Electronics Engineers (IEEE)},
}

@Article{Angerbauer2023,
  author    = {Angerbauer, Stefan and others},
  journal   = {IEEE Trans. Mol. Biol. Multi-Scale Commun.},
  title     = {Salinity-based molecular communication in microfluidic channels},
  year      = {2023},
  issn      = {2332-7804},
  month     = jun,
  number    = {2},
  pages     = {191--206},
  volume    = {9},
  doi       = {10.1109/tmbmc.2023.3277391},
  publisher = {Institute of Electrical and Electronics Engineers (IEEE)},
}

@InProceedings{Huang2024,
  author    = {Huang, Yu and Huang, Xuewei and Ji, Fei and Cheng, Mingyue and Chen, Xuan and Wen, Miaowen},
  booktitle = {IEEE Conf. Comput. Commun. Works.},
  title     = {Demo: A non-invasive and high-speed molecular communication testbed with capacitive sensing},
  year      = {2024},
  month     = may,
  pages     = {1--2},
  doi       = {10.1109/infocomwkshps61880.2024.10620884},
}

@InProceedings{Brand2024,
  author    = {Brand, Lukas and others},
  booktitle = {Proc. IEEE Int. Conf. Commun.},
  title     = {Closed loop molecular communication testbed: setup, interference analysis, and experimental results},
  year      = {2024},
  month     = jun,
  pages     = {4805--4811},
  doi       = {10.1109/icc51166.2024.10622231},
}

@InProceedings{Koo2020,
  author    = {Koo, Bon-Hong and Kim, Ho Joong and Kwon, Jang-Yeon and Chae, Chan-Byoung},
  booktitle = {Proc. IEEE Int. Conf. Commun.},
  title     = {Deep learning-based human implantable nano molecular communications},
  year      = {2020},
  month     = jun,
  pages     = {1--7},
  doi       = {10.1109/icc40277.2020.9148818},
}

@InProceedings{Farsad2017,
  author    = {Farsad, Nariman and Pan, David and Goldsmith, Andrea},
  booktitle = {Proc. IEEE Glob. Commun. Conf.},
  title     = {A novel experimental platform for in-vessel multi-chemical molecular communications},
  year      = {2017},
  month     = dec,
  doi       = {10.1109/glocom.2017.8255058},
}

@Article{Pan2022,
  author    = {Pan, Wenxin and Chen, Xiaokang and Yang, Xiaodong and Zhao, Nan and Meng, Lingguo and Shah, Fiaz Hussain},
  journal   = {Nanomaterials},
  title     = {A molecular communication platform based on body area nanonetwork},
  year      = {2022},
  issn      = {2079-4991},
  month     = feb,
  number    = {4},
  pages     = {722},
  volume    = {12},
  doi       = {10.3390/nano12040722},
  publisher = {MDPI AG},
}

@Article{Tuccitto2017,
  author  = {Tuccitto, Nunzio and Li-Destri, Giovanni and Messina, Grazia M. L. and Marletta, Giovanni},
  journal = {J. Phys. Chem. Lett.},
  title   = {Fluorescent quantum dots make feasible long-range transmission of molecular bits},
  year    = {2017},
  month   = aug,
  number  = {16},
  pages   = {3861--3866},
  volume  = {8},
  doi     = {10.1021/acs.jpclett.7b01713},
}

@Article{Lin2024,
  author    = {Lin, Lin and Wang, Wei and Yu, Wenlong and Yan, Hao},
  journal   = {IEEE Trans. Mol. Biol. Multi-Scale Commun.},
  title     = {Testbed for molecular communication system based on light absorption: Study of information transmission from inside to outside body},
  year      = {2024},
  issn      = {2332-7804},
  month     = jun,
  number    = {2},
  pages     = {212--222},
  volume    = {10},
  doi       = {10.1109/tmbmc.2024.3379282},
  publisher = {Institute of Electrical and Electronics Engineers (IEEE)},
}

@Article{Hamidovic2024,
  author    = {Hamidović, Medina and Angerbauer, Stefan and Bi, Dadi and Deng, Yansha and Tugcu, Tuna and Haselmayr, Werner},
  journal   = {IEEE Trans. Mol. Biol. Multi-Scale Commun.},
  title     = {Microfluidic systems for molecular communications: A review from theory to practice},
  year      = {2024},
  issn      = {2332-7804},
  month     = mar,
  number    = {1},
  pages     = {147--163},
  volume    = {10},
  doi       = {10.1109/tmbmc.2024.3368768},
  publisher = {Institute of Electrical and Electronics Engineers (IEEE)},
}

@Article{Chahibi2017,
  author    = {Chahibi, Youssef},
  journal   = {Nano Commun. Netw.},
  title     = {Molecular communication for drug delivery systems: A survey},
  year      = {2017},
  issn      = {1878-7789},
  month     = mar,
  pages     = {90--102},
  volume    = {11},
  doi       = {10.1016/j.nancom.2017.01.003},
  publisher = {Elsevier BV},
}

@Article{ChudeOkonkwo2017,
  author    = {Chude-Okonkwo, Uche A. K. and Malekian, Reza and Maharaj, B. T. and Vasilakos, Athanasios V.},
  journal   = {IEEE Commun. Surv. Tutor.},
  title     = {Molecular communication and nanonetwork for targeted drug delivery: A survey},
  year      = {2017},
  issn      = {1553-877X},
  month     = may,
  number    = {4},
  pages     = {3046--3096},
  volume    = {19},
  doi       = {10.1109/comst.2017.2705740},
  publisher = {Institute of Electrical and Electronics Engineers (IEEE)},
}

@InProceedings{Wang2020,
  author     = {Wang, Jiaming and Hu, Dongyin and Shetty, Chirag and Hassanieh, Haitham},
  booktitle  = {Proc. Int. Conf. Mob. Comput. Netw.},
  title      = {Understanding and embracing the complexities of the molecular communication channel in liquids},
  year       = {2020},
  month      = sep,
  pages      = {1--15},
  collection = {MobiCom ’20},
  doi        = {10.1145/3372224.3419191},
}

@Article{Etemadi2023,
  author    = {Etemadi, Ali and Farahnak-Ghazani, Maryam and Arjmandi, Hamidreza and Mirmohseni, Mahtab and Nasiri-Kenari, Masoumeh},
  journal   = {IEEE Access},
  title     = {Abnormality detection and localization schemes using molecular communication systems: A survey},
  year      = {2023},
  issn      = {2169-3536},
  pages     = {1761--1792},
  volume    = {11},
  doi       = {10.1109/access.2022.3228618},
  publisher = {Institute of Electrical and Electronics Engineers (IEEE)},
}

@Article{Sun2022,
  author    = {Sun, Yue and Liu, Meiling and Xiao, Yue and Chen, Yifan},
  journal   = {Comput. Methods Programs Biomed.},
  title     = {A novel molecular communication inspired detection method for the evolution of atherosclerosis},
  year      = {2022},
  issn      = {0169-2607},
  month     = jun,
  pages     = {106756},
  volume    = {219},
  doi       = {10.1016/j.cmpb.2022.106756},
  publisher = {Elsevier BV},
}

@Article{Khaloopour2021,
  author    = {Khaloopour, Ladan and Mirmohseni, Mahtab and Nasiri-Kenari, Masoumeh},
  journal   = {IEEE Sens. J.},
  title     = {Theoretical concept study of cooperative abnormality detection and localization in fluidic-medium molecular communication},
  year      = {2021},
  issn      = {2379-9153},
  month     = aug,
  number    = {15},
  pages     = {17118--17130},
  volume    = {21},
  doi       = {10.1109/jsen.2021.3081815},
  publisher = {Institute of Electrical and Electronics Engineers (IEEE)},
}

@Article{Zafar2021,
  author    = {Zafar, Sidra and others},
  journal   = {IEEE Access},
  title     = {A systematic review of bio-cyber interface technologies and security issues for {I}nternet of {B}io-{N}ano {T}hings},
  year      = {2021},
  issn      = {2169-3536},
  month     = jun,
  pages     = {93529--93566},
  volume    = {9},
  doi       = {10.1109/access.2021.3093442},
  publisher = {Institute of Electrical and Electronics Engineers (IEEE)},
}

@Article{Kuscu2021,
  author    = {Murat Kuscu and Bige Deniz Unluturk},
  journal   = {ITU J. Future Evol. Technol.},
  title     = {Internet of {B}io-{N}ano {T}hings: A review of applications, enabling technologies and key challenges},
  year      = {2021},
  issn      = {2616-8375},
  month     = dec,
  number    = {3},
  pages     = {1--24},
  volume    = {2},
  doi       = {10.52953/chbb9821},
  publisher = {International Telecommunication Union},
}

@Article{Schafer2021,
  author    = {Sch\"afer, Maximilian and Wicke, Wayan and Brand, Lukas and Rabenstein, Rudolf and Schober, Robert},
  journal   = {IEEE Trans. Mol. Biol. Multi-Scale Commun.},
  title     = {Transfer function models for cylindrical {MC} channels with diffusion and laminar flow},
  year      = {2021},
  issn      = {2332-7804},
  month     = dec,
  number    = {4},
  pages     = {271--287},
  volume    = {7},
  doi       = {10.1109/tmbmc.2021.3061030},
  publisher = {Institute of Electrical and Electronics Engineers (IEEE)},
}

@InProceedings{Noel2014a,
  author    = {Noel, Adam and Cheung, Karen C. and Schober, Robert},
  booktitle = {Proc. IEEE Int. Conf. Commun.},
  title     = {Diffusive molecular communication with disruptive flows},
  year      = {2014},
  month     = jun,
  pages     = {3600--3606},
  doi       = {10.1109/icc.2014.6883880},
}

@Article{Noel2014b,
  author    = {Noel, Adam and Cheung, Karen C. and Schober, Robert},
  journal   = {IEEE Trans. Nanobiosci.},
  title     = {Optimal receiver design for diffusive molecular communication with flow and additive noise},
  year      = {2014},
  issn      = {1558-2639},
  month     = sep,
  number    = {3},
  pages     = {350--362},
  volume    = {13},
  doi       = {10.1109/tnb.2014.2337239},
  publisher = {Institute of Electrical and Electronics Engineers (IEEE)},
}

@Book{Aaronson2012,
  author    = {Aaronson, Philip I. and Ward, Jeremy P. T. and Connolly, Michelle J.},
  publisher = {Wiley \& Sons, Incorporated, John},
  title     = {The {C}ardiovascular {S}ystem at a {G}lance},
  year      = {2012},
  isbn      = {9781118535400},
  pages     = {136},
}

@InProceedings{Wicke2018,
  author    = {Wicke, Wayan and Schwering, Tobias and Ahmadzadeh, Arman and Jamali, Vahid and Noel, Adam and Schober, Robert},
  booktitle = {Proc. IEEE Glob. Commun. Conf.},
  title     = {Modeling duct flow for molecular communication},
  year      = {2018},
  month     = dec,
  pages     = {206--212},
  doi       = {10.1109/glocom.2018.8647632},
}

@Article{Dhok2022,
  author    = {Dhok, Shivani and Chouhan, Lokendra and Noel, Adam and Sharma, Prabhat Kumar},
  journal   = {IEEE Trans. Mol. Biol. Multi-Scale Commun.},
  title     = {Cooperative molecular communication in drift-induced diffusive cylindrical channel},
  year      = {2022},
  issn      = {2332-7804},
  month     = mar,
  number    = {1},
  pages     = {44--55},
  volume    = {8},
  doi       = {10.1109/tmbmc.2021.3089939},
  publisher = {Institute of Electrical and Electronics Engineers (IEEE)},
}

@Article{Varshney2019,
  author    = {Varshney, Neeraj and Patel, Adarsh and Haselmayr, Werner and Jagannatham, Aditya K. and Varshney, Pramod K. and Nallanathan, Arumugam},
  journal   = {IEEE Trans. Commun.},
  title     = {Impact of intermediate nanomachines in multiple cooperative nanomachine-assisted diffusion advection mobile molecular communication},
  year      = {2019},
  issn      = {1558-0857},
  month     = jul,
  number    = {7},
  pages     = {4856--4871},
  volume    = {67},
  doi       = {10.1109/tcomm.2019.2909900},
  publisher = {Institute of Electrical and Electronics Engineers (IEEE)},
}

@Article{Farsad2016,
  author    = {Farsad, Nariman and Yilmaz, H. Birkan and Eckford, Andrew and Chae, Chan-Byoung and Guo, Weisi},
  journal   = {IEEE Commun. Surv. Tutor.},
  title     = {A comprehensive survey of recent advancements in molecular communication},
  year      = {2016},
  issn      = {1553-877X},
  month     = feb,
  number    = {3},
  pages     = {1887--1919},
  volume    = {18},
  doi       = {10.1109/comst.2016.2527741},
  publisher = {Institute of Electrical and Electronics Engineers (IEEE)},
}

@Article{Jamali2019,
  author    = {Jamali, Vahid and Ahmadzadeh, Arman and Wicke, Wayan and Noel, Adam and Schober, Robert},
  journal   = {Proc. IEEE},
  title     = {Channel modeling for diffusive molecular communication—{A} tutorial review},
  year      = {2019},
  issn      = {1558-2256},
  month     = jul,
  number    = {7},
  pages     = {1256--1301},
  volume    = {107},
  doi       = {10.1109/jproc.2019.2919455},
  publisher = {Institute of Electrical and Electronics Engineers (IEEE)},
}

@Article{Bicen2013,
  author    = {Bicen, A. Ozan and Akyildiz, Ian F.},
  journal   = {IEEE Trans. Signal Process.},
  title     = {System-theoretic analysis and least-squares design of microfluidic channels for flow-induced molecular communication},
  year      = {2013},
  issn      = {1941-0476},
  month     = oct,
  number    = {20},
  pages     = {5000--5013},
  volume    = {61},
  doi       = {10.1109/tsp.2013.2274959},
  publisher = {Institute of Electrical and Electronics Engineers (IEEE)},
}

@InProceedings{Schafer2019,
  author    = {Sch\"afer, Maximilian and Wicke, Wayan and Rabenstein, Rudolf and Schober, Robert},
  booktitle = {Proc. IEEE Int. Conf. Commun.},
  title     = {Analytical models for particle diffusion and flow in a horizontal cylinder with a vertical force},
  year      = {2019},
  month     = may,
  pages     = {1--7},
  doi       = {10.1109/icc.2019.8761148},
}

@Article{Chahibi2015,
  author    = {Chahibi, Youssef and Pierobon, Massimiliano and Akyildiz, Ian F.},
  journal   = {IEEE Trans. Biomed. Eng.},
  title     = {Pharmacokinetic modeling and biodistribution estimation through the molecular communication paradigm},
  year      = {2015},
  issn      = {1558-2531},
  month     = oct,
  number    = {10},
  pages     = {2410--2420},
  volume    = {62},
  doi       = {10.1109/tbme.2015.2430011},
  publisher = {Institute of Electrical and Electronics Engineers (IEEE)},
}

@Article{Lo2019,
  author    = {Lo, Yun-Feng and Lee, Chia-Han and Chou, Po-Chun and Yeh, Ping-Cheng},
  journal   = {IEEE Commun. Lett.},
  title     = {Modeling molecular communications in tubes with poiseuille flow and robin boundary condition},
  year      = {2019},
  issn      = {2373-7891},
  month     = aug,
  number    = {8},
  pages     = {1314--1318},
  volume    = {23},
  doi       = {10.1109/lcomm.2019.2920830},
  publisher = {Institute of Electrical and Electronics Engineers (IEEE)},
}

@Article{Kuscu2018,
  author    = {Kuscu, Murat and Akan, Ozgur B.},
  journal   = {PLOS ONE},
  title     = {Modeling convection-diffusion-reaction systems for microfluidic molecular communications with surface-based receivers in {I}nternet of {B}io-{N}ano {T}hings},
  year      = {2018},
  issn      = {1932-6203},
  month     = feb,
  number    = {2},
  pages     = {e0192202},
  volume    = {13},
  doi       = {10.1371/journal.pone.0192202},
  editor    = {Liu, Pan-Ping},
  publisher = {Public Library of Science (PLoS)},
}

@Article{Chahibi2015a,
  author    = {Chahibi, Youssef and Akyildiz, Ian F. and Balasubramaniam, Sasitharan and Koucheryavy, Yevgeni},
  journal   = {IEEE Trans. Biomed. Eng.},
  title     = {Molecular communication modeling of antibody-mediated drug delivery systems},
  year      = {2015},
  issn      = {1558-2531},
  month     = jul,
  number    = {7},
  pages     = {1683--1695},
  volume    = {62},
  doi       = {10.1109/tbme.2015.2400631},
  publisher = {Institute of Electrical and Electronics Engineers (IEEE)},
}

@Article{Jiang2022,
  author    = {Jiang, Weiquan and Zeng, Li and Fu, Xudong and Wu, Zi},
  journal   = {J. Fluid Mech.},
  title     = {Analytical solutions for reactive shear dispersion with boundary adsorption and desorption},
  year      = {2022},
  issn      = {1469-7645},
  month     = aug,
  volume    = {947},
  doi       = {10.1017/jfm.2022.656},
  publisher = {Cambridge University Press (CUP)},
}

@Article{Aris1956,
  author    = {R. Aris},
  journal   = {Proc. R. Soc.},
  title     = {On the dispersion of a solute in a fluid flowing through a tube},
  year      = {1956},
  issn      = {2053-9169},
  month     = apr,
  pages     = {67--77},
  volume    = {235},
  doi       = {10.1098/rspa.1956.0065},
  publisher = {The Royal Society},
}

@Article{Jamali2023,
  author    = {Jamali, Vahid and Loos, Helene M. and Buettner, Andrea and Schober, Robert and Vincent Poor, H.},
  journal   = {IEEE Trans. Commun.},
  title     = {Olfaction-inspired {MC}s: {M}olecule mixture shift keying and cross-reactive receptor arrays},
  year      = {2023},
  issn      = {1558-0857},
  month     = apr,
  number    = {4},
  pages     = {1894--1911},
  volume    = {71},
  doi       = {10.1109/tcomm.2023.3242379},
  publisher = {Institute of Electrical and Electronics Engineers (IEEE)},
}

@Article{Felicetti2016,
  author    = {Felicetti, L. and Femminella, M. and Reali, G. and Liò, P.},
  journal   = {Nano Commun. Netw.},
  title     = {Applications of molecular communications to medicine: A survey},
  year      = {2016},
  issn      = {1878-7789},
  month     = mar,
  pages     = {27--45},
  volume    = {7},
  doi       = {10.1016/j.nancom.2015.08.004},
  publisher = {Elsevier BV},
}

@Article{Giddings1955,
  author    = {Giddings, J. Calvin and Eyring, Henry},
  journal   = {J. Phys. Chem.},
  title     = {A molecular dynamic theory of chromatography},
  year      = {1955},
  issn      = {1541-5740},
  month     = may,
  number    = {5},
  pages     = {416--421},
  volume    = {59},
  doi       = {10.1021/j150527a009},
  publisher = {American Chemical Society (ACS)},
}

@InProceedings{Jakumeit2025,
  author    = {Jakumeit, Timo and Brand, Lukas and Kirchner, Jens and Schober, Robert and Lotter, Sebastian},
  booktitle = {Proc. IEEE Glob. Commun. Conf.},
  title     = {Molecular signal reception in complex vessel networks: The role of the network topology},
  year      = {2025},
  month     = jun,
  pages     = {6049--6055},
  doi       = {10.1109/icc52391.2025.11160711},
}

@InProceedings{Wietfeld2024,
  author    = {Wietfeld, Alexander and Schmidt, Sebastian and Kellerer, Wolfgang},
  booktitle = {Proc. IEEE Glob. Commun. Conf.},
  title     = {Evaluation of a multi-molecule molecular communication testbed based on pectral sensing},
  year      = {2024},
  month     = dec,
  pages     = {2102--2108},
  doi       = {10.1109/globecom52923.2024.10901414},
}

@Article{Vakilipoor2025,
  author    = {Vakilipoor, Fardad and others},
  journal   = {IEEE Trans. Mol. Biol. Multi-Scale Commun.},
  title     = {The {CAM} model: {A}n in vivo testbed for molecular communication systems},
  year      = {2025},
  issn      = {2332-7804},
  month     = aug,
  pages     = {1--1},
  doi       = {10.1109/tmbmc.2025.3601432},
  publisher = {Institute of Electrical and Electronics Engineers (IEEE)},
}

@Article{Gao2010,
  author    = {Gao, Weiwei and Chan, Juliana M. and Farokhzad, Omid C.},
  journal   = {Mol. Pharm.},
  title     = {p{H}-responsive nanoparticles for drug delivery},
  year      = {2010},
  issn      = {1543-8392},
  month     = oct,
  number    = {6},
  pages     = {1913--1920},
  volume    = {7},
  doi       = {10.1021/mp100253e},
  publisher = {American Chemical Society (ACS)},
}

@Article{Lotter2023a,
  author    = {Lotter, Sebastian and others},
  journal   = {IEEE Nanotechnol. Mag.},
  title     = {Experimental research in synthetic molecular communications – {P}art {II}},
  year      = {2023},
  issn      = {1942-7808},
  month     = jun,
  number    = {3},
  pages     = {54--65},
  volume    = {17},
  doi       = {10.1109/mnano.2023.3262377},
  publisher = {Institute of Electrical and Electronics Engineers (IEEE)},
}

@misc{jakumeit2025vessel,
  author       = {T. Jakumeit and L. Brand and J. Kirchner and R. Schober and S. Lotter},
  title        = {{Vessel Network Topology in Molecular Communication: Insights from Experiments and Theory}},
  howpublished = {Zenodo},
  year         = {2025},
  month        = nov,
  day          = {11},
  doi          = {10.5281/zenodo.17581905},
  url          = {https://doi.org/10.5281/zenodo.17581905}
}

\end{document}